\documentclass[numbers,sort&compress]{article}
\usepackage[utf8]{inputenc}
\usepackage[T1]{fontenc}
\usepackage{graphicx}
\usepackage{grffile}
\usepackage{longtable}
\usepackage{wrapfig}
\usepackage{rotating}
\usepackage[normalem]{ulem}
\usepackage{amsmath}
\usepackage{amssymb}
\usepackage{mathtools}
\usepackage{textcomp}
\usepackage{capt-of}
\usepackage{hyperref}
\usepackage{arxiv}
\usepackage{amssymb,amsmath,amsthm}
\usepackage[square,sort,comma,numbers]{natbib}
\usepackage{abbrevs}
\usepackage{hyperref}
\usepackage{algorithm2e}
\theoremstyle{definition}
\usepackage{url}            %
\usepackage{booktabs}       %
\usepackage{amsfonts}       %
\usepackage{nicefrac}       %
\usepackage{microtype}      %
\usepackage{cleveref}       %
\usepackage{graphicx}
\usepackage{doi}

\graphicspath{{src/}}

\usepackage{framed,multirow}
\usepackage{url}
\usepackage[usenames,dvipsnames]{xcolor}
\definecolor{newcolor}{rgb}{.8,.349,.1}

\newcounter{bla}

\usepackage{amssymb}
\usepackage{amsmath}
\usepackage{physics}
\usepackage{braket}
\usepackage{cancel}
\usepackage[utf8]{inputenc}
\usepackage{subfig}
\usepackage{booktabs}
\usepackage{tablefootnote}

\newcommand{\inrm}{\mathrm{in}}
\newcommand{\outrm}{\mathrm{out}}

\newcommand{\half}{{\textstyle\frac{1}{2}}}

\author{
\href{https://orcid.org/0000-0003-3968-3614}{\includegraphics[scale=0.06]{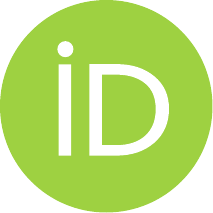}\hspace{1mm}Ondřej Čertík}\thanks{Currently at GSI Technology, USA} \\
Computational Physics and Methods (CCS-2)\\
Los Alamos National Laboratory\\
NM 87545, USA\\
\texttt{ondrej@certik.us}
\And
John E. Pask \\
Physics Division\\
Lawrence Livermore National Laboratory\\
Livermore, CA 94550, USA\\
\texttt{pask1@llnl.gov}
\And
\href{https://orcid.org/0000-0002-1126-5448}{\includegraphics[scale=0.06]{orcid.pdf}\hspace{1mm}Isuru Fernando} \\
Department of Computer Science\\
University of Illinois at Urbana--Champaign\\
Urbana, IL 61801, USA\
\texttt{isuruf@gmail.com}
\And
\href{https://orcid.org/0000-0002-2393-8056}{\includegraphics[scale=0.06]{orcid.pdf}\hspace{1mm}Rohit Goswami}\thanks{Corresponding Author} \\
Science Institute, University of Iceland\\
Quansight Labs, TX, Austin\\
\texttt{rgoswami@ieee.org}
\And
N. Sukumar \\
Department of Civil and Environmental Engineering\\
University of California, Davis\\
CA 95616, USA\\
\texttt{nsukumar@ucdavis.edu}
\And
\href{https://orcid.org/0000-0002-8180-9367}{\includegraphics[scale=0.06]{orcid.pdf}\hspace{1mm}Lee A. Collins} \\
Physics and Chemistry of Materials (T-1)\\
Los Alamos National Laboratory\\
NM 87545, USA\\
\texttt{lac@lanl.gov}
\And
Gianmarco Manzini \\
Applied Mathematics and Plasma Physics (T-5)\\
Los Alamos National Laboratory\\
NM 87545, USA\\
\texttt{gmanzini@lanl.gov}
\And
\href{https://orcid.org/0000-0002-4491-6841}{\includegraphics[scale=0.06]{orcid.pdf}\hspace{1mm}Jiří Vackář} \\
Institute of Physics\\
Academy of Sciences of the Czech Republic\\
Na Slovance 2, 182 21 Praha 8, Czech Republic\\
\texttt{vackar@fzu.cz}
}

\date{\today}
\title{High-order finite element method for atomic structure calculations}
\hypersetup{
 pdfauthor={Ondrej, John, Isuru, Rohit, Sukumar, lee, Gianmarco, Jiri},
 pdftitle={High-order finite element method for atomic structure calculations},
 pdfkeywords={atomic structure, electronic structure, Schrödinger equation, Dirac equation, Kohn-Sham equations, density functional theory, finite element method, Fortran 2008},
 pdfsubject={},
 pdfcreator={Emacs 28.2.50},
 pdflang={English}}
\begin{document}

\maketitle

\begin{abstract}
We introduce \texttt{featom}, an open source code that implements a high-order
finite element solver for the radial Schrödinger, Dirac, and Kohn-Sham
equations. The formulation accommodates various mesh types, such as uniform or
exponential, and the convergence can be systematically controlled by increasing
the number and/or polynomial order of the finite element basis functions. The
Dirac equation is solved using a squared Hamiltonian approach to eliminate
spurious states. To address the slow convergence of the $\kappa=\pm1$ states due to
divergent derivatives at the origin, we incorporate known asymptotic forms into
the solutions. We achieve a high level of accuracy ($10^{-8}$ Hartree) for total
energies and eigenvalues of heavy atoms such as uranium in both Schrödinger and
Dirac Kohn-Sham solutions. We provide detailed convergence studies and
computational parameters required to attain commonly required accuracies.
Finally, we compare our results with known analytic results as well as the
results of other methods. In particular, we calculate benchmark results for
atomic numbers ($Z$) from 1 to 92, verifying current benchmarks. We demonstrate
significant speedup compared to the state-of-the-art shooting solver
\texttt{dftatom}.  An efficient, modular Fortran 2008 implementation, is
provided under an open source, permissive license, including examples and tests,
wherein particular emphasis is placed on the independence (no global variables),
reusability, and generality of the individual routines.
\end{abstract}
\keywords{atomic structure, electronic structure, Schrödinger equation, Dirac equation, Kohn-Sham equations, density functional theory, finite element method, Fortran 2008}

\section{Introduction}\label{introduction}

Over the past three decades, Density Functional Theory (DFT)
\cite{hohenbergInhomogeneousElectronGas1964} has established itself as a
cornerstone of modern materials research, enabling the understanding,
prediction, and control of a wide variety of materials properties from the first
principles of quantum mechanics, with no empirical parameters.  However, the
solution of the required Kohn-Sham equations
\cite{kohnSelfConsistentEquationsIncluding1965} is a formidable task, which has
given rise to a number of different solution methods \cite{martinElectronicStructureBasic2004}. At the heart of
the majority of methods in use today, whether for isolated systems such as
molecules or extended systems such as solids and liquids, lies the solution of
the Schrödinger and/or Dirac equations for the isolated atoms composing the
larger molecular or condensed matter systems of interest. Particular challenges
arise in the context of relativistic calculations, which require solving the
Dirac equation, since spurious states can arise due to the unbounded nature of
the Dirac Hamiltonian operator and inconsistencies in discretizations derived
therefrom \cite{grantRelativisticQuantumTheory2007}.

A number of approaches have been developed to avoid spurious states in the
solution of the Dirac equation. Shooting methods, e.g.,
\cite{certikDftatomRobustGeneral2013} and references therein, avoid such states
by leveraging known asymptotic forms to target desired eigenfunctions based on
selected energies and numbers of nodes. However, due to the need for many trial
solutions to find each eigenfunction, efficient implementation while maintaining
robustness is nontrivial. In addition, convergence parameters such as distance
of grid points from the origin must be carefully tuned. Basis set methods, e.g.,
\cite{dyallKineticBalanceVariational1990,fischerBsplineGalerkinMethod2009,grantBsplineMethodsRadial2009,almanasrehStabilizedFiniteElement2013}
and references therein, offer an elegant alternative to shooting methods,
solving for all states at once by diagonalization of the Schrödinger or Dirac
Hamiltonian in the chosen basis. However, due to the unbounded spectrum of the
Dirac Hamiltonian, spurious states have been a longstanding issue
\cite{grantRelativisticQuantumTheory2007,tupitsynSpuriousStatesDirac2008}. Many approaches have been developed to avoid spurious states over the past few decades, with varying degrees of
success. These include using different bases for the large and small components
of the Dirac wavefunction
\cite{dyallKineticBalanceVariational1990,shabaevDualKineticBalance2004,beloyApplicationDualkineticbalanceSets2008,fischerBsplineGalerkinMethod2009,sunComparisonRestrictedUnrestricted2011,jiaoDevelopmentKineticallyAtomically2021},
modifying the Hamiltonian
\cite{kutzelniggBasisSetExpansion1984,grantBsplineMethodsRadial2009,almanasrehStabilizedFiniteElement2013,almanasrehFiniteElementMethod2019,fangSolutionDiracEquation2020},
and imposing various boundary constraints
\cite{johnsonFiniteBasisSets1988,sapirsteinUseBasisSplines1996,beloyApplicationDualkineticbalanceSets2008,grantBsplineMethodsRadial2009}.
In the finite-difference context, defining large and small components of the
Dirac wavefunction on alternate grid points
\cite{salomonsonRelativisticAllorderPair1989}, replacing conventional central
differences with asymmetric differences
\cite{salomonsonRelativisticAllorderPair1989,zhangResolvingSpuriousstateProblem2022},
and adding a Wilson term to the Hamiltonian \cite{fangSolutionDiracEquation2020}
have proven effective in eliminating spurious states. While in the finite
element (FE) context, the use of different trial and test spaces in a stabilized
Petrov-Galerkin formulation has proven effective in mitigating the large
off-diagonal convection (first derivative) terms and absence of diffusion
(second derivative) terms causing the instability
\cite{almanasrehStabilizedFiniteElement2013,almanasrehFiniteElementMethod2019}.

In the this work, we present the open-source code,\texttt{featom}\footnote{Source also publicly on Github: \url{https://github.com/atomic-solvers/featom}}, for the
solution of the Schrödinger, Dirac, and Kohn-Sham equations in a high-order
finite element basis. The FE basis enables exponential convergence with respect
to polynomial order while allowing full flexibility as to choice of radial mesh.
To eliminate spurious states in the solution of the Dirac equation, we square
the Dirac Hamiltonian operator
\cite{wallmeierUseSquaredDirac1981,kutzelniggBasisSetExpansion1984}. Since the
square of the operator has the same eigenfunctions as the operator itself, and
the square of its eigenvalues, determination of the desired eigenfunctions and
eigenvalues is immediate. Most importantly, since the square of the operator is
bounded from below, unlike the operator itself, it is amenable to direct
solution by standard variational methods, such as FE, without modification. This
affords simplicity, robustness, and well understood convergence. Moreover,
squaring the operator rather than modifying it and/or boundary conditions upon
it, ensures key properties are preserved exactly, such as convergence to the
correct non-relativistic (Schrödinger) limit with increasing speed of light
\cite{kutzelniggBasisSetExpansion1984}. Squaring the operator also stabilizes
the numerics naturally, without approximation, by creating second-derivative
terms. To accelerate convergence with respect to polynomial order, we
incorporate known asymptotic forms as $r\to0$ into the solutions: rather than
solving for large and small Dirac wavefunction components $P(r)$ and $Q(r)$, we
solve for $\frac{P(r)}{r^\alpha}$ and $\frac{Q(r)}{r^\alpha}$, with $\alpha$ based on
the known asymptotic forms for $P$ and $Q$ as $r\to0$. This eliminates derivative divergences and
non-polynomial behavior in the vicinity of the origin and so enables rapidly
convergent solutions in a polynomial basis for all quantum numbers $\kappa$,
including $\kappa=\pm1$. By combining the above ideas, \texttt{featom} is able to
provide robust, efficient, and accurate solutions for both Schrödinger and Dirac
equations.

The package is MIT licensed and is written in Fortran, leveraging language
features from the 2008 standard, with an emphasis on facilitating user
extensions. Additionally, it is designed to work within the modern Fortran
ecosystem and leverages the \texttt{fpm} build system. There are several
benchmark calculations that serve as tests. The package supports different
mesh-generating methods, including support for a uniform mesh, an exponential
mesh, and other meshes defined by nodal distributions and derivatives.
Multiple quadrature methods have been implemented, and their
usage in the code is physically motivated. Gauss--Jacobi
quadrature is used to accurately integrate problematic integrals for the Dirac
equation and the Poisson equation as well as for the total energy in the
Dirac-Kohn-Sham solution, whereas Gauss--Legendre quadrature is used for the Schrödinger
equation. To ensure numerical accuracy, we employ several techniques, using
Gauss--Lobatto quadrature for the overlap matrix to recover a standard eigenproblem,
precalculation of most quantities, and parsimonious assembly of a lower-triangular matrix with a symmetric eigensolver. With these considerations we
show that the resulting code outperforms the state-of-the-art code \texttt{dftatom}, using
fewer parameters for convergence while retaining high accuracy of $10^{-8}$ Hartree
in total energy and eigenvalues for uranium (both Dirac and Schrödinger) and
all lighter atoms.

The remainder of the paper is organized as follows. Section~\ref{escbasic}
describes the electronic structure equations solved. This is followed by
Sections~\ref{ferschrod} and \ref{ferdirac}, which detail the unified finite
element solution for the radial Schrödinger and Dirac equations.
Section~\ref{numericaleff} details the numerical techniques employed to
efficiently construct and solve the resulting matrix eigenvalue problem,
including mesh and quadrature methods. In Section~\ref{results}, we present
results from analytic tests and benchmark comparisons against the shooting
solver \texttt{dftatom}, followed by a brief discussion of findings.
Finally, in Section~\ref{conclusions}, we summarize our main conclusions.

\section{Electronic structure equations}\label{escbasic}
Under the assumption of a central potential, we establish the conventions used
for the electronic structure problems under the purview of \texttt{featom} in
this section. Starting from the nonrelativistic radial Schrödinger equation and
its relativistic counterpart, the Dirac equation, we couple these to the Kohn--Sham
equations with a Poisson equation for the Hartree potential. The interested reader may find more details of this formulation in
\cite{certikDftatomRobustGeneral2013}. By convention, we use Hartree atomic
units throughout the manuscript.

\subsection{Radial Schrödinger equation}\label{rschrod}
Recall that the 3D one-electron Schrödinger equation is given by
\begin{equation}
\left(-\half\nabla^2+V({\bf x})\right)\psi({\bf x}) = E\psi({\bf x}).
\end{equation}

\noindent When the potential considered is spherically symmetric, i.e.,
\begin{equation}
V({\bf x}):=V(r),
\end{equation}
the eigenstates of energy and angular momentum can be written in the form
\begin{equation}
\psi_{nlm}({\bf x})=R_{nl}(r)\,Y_{lm}\left(\frac{{\bf x}}{r}\right),
\label{psi}
\end{equation}
where $n$ is the principal quantum number, $l$ is the orbital angular momentum
quantum number, and $m$ is the magnetic quantum number.
It follows that $R_{nl}(r)$ satisfies the radial Schrödinger equation
\begin{equation}
-\half\left(r^2 R_{nl}'(r)\right)' + \left(r^2 V + \half l(l+1)\right)R_{nl}(r)
    =E r^2 R_{nl}(r).
\label{radial}
\end{equation}

\noindent The functions $\psi_{nlm}({\bf x})$ and $R_{nl}(r)$ are normalized as
\begin{align}
\int |\psi_{nlm}({\bf x})|^2 \dd[3]{x} &= 1, \nonumber \\
\int_0^\infty R_{nl}^2(r) r^2 \dd r &= 1.
\end{align}

\subsection{Radial Dirac equation}\label{rdirac}
The one-electron radial Dirac equation can be written as
\begin{subequations}
\begin{align}
\label{dirac_eq1}
P_{n\kappa}'(r) &= -{\frac{\kappa}{r}}P_{n\kappa}(r)+\left({\frac{E-V(r)}{c}}+2c\right)Q_{n\kappa}(r), \\
\label{dirac_eq2}
Q_{n\kappa}'(r) &= -\left({E-\frac{V(r)}{c}}\right)P_{n\kappa}(r)+{\frac{\kappa}{r}}Q_{n\kappa}(r),
\end{align}
\end{subequations}
where $P_{n\kappa}(r)$ and $Q_{n\kappa}(r)$ are related to the usual large $g_{n\kappa}(r)$ and small $f_{n\kappa}(r)$ components
of the Dirac equation by
\begin{subequations}
\begin{align}
P_{n\kappa}(r)&=rg_{n\kappa}(r), \\
Q_{n\kappa}(r)&=rf_{n\kappa}(r).
\end{align}
\end{subequations}
A pedagogical derivation of these results can be found in the literature, for
example
in~\cite{certikDftatomRobustGeneral2013,strangeRelativisticQuantumMechanics1998,zabloudilElectronScatteringSolid2006}.
We follow the solution labeling in \cite{certikDftatomRobustGeneral2013}, in
which the relativistic quantum number $\kappa$ is determined by the orbital
angular momentum quantum number $l$ and spin quantum number $s=\pm1$ on the basis
of the total angular momentum quantum number $j=l\pm\frac{1}{2}$ using
\begin{equation}
\kappa    = \begin{cases}
                -l-1 & \text{for $j=l+\half$, i.e. $s=+1$}, \\
                \quad l   & \text{for $j=l-\half$, i.e. $s=-1$}.
            \end{cases}
\end{equation}
By not including the rest mass energy of an electron, the energies obtained
from the radial Dirac equation can be compared to the non-relativistic energies obtained from the Schrödinger equation.

\noindent The normalization of $P_{n\kappa}(r)$ and $Q_{n\kappa}(r)$ is
\begin{equation}
\int_0^\infty (P_{n\kappa}^2(r) + Q_{n\kappa}^2(r)) \dd r = 1.
\end{equation}

\noindent We note that both $P_{n\kappa}$ and $Q_{n\kappa}$ are solutions of
homogeneous equations and are thus only unique up to an arbitrary multiplicative
constant.

\subsection{Poisson equation}\label{poisson}
The 3D Poisson equation for the Hartree potential $V_H$
due to electronic density $n$ is given by
\begin{equation}
\nabla^2V_H({\bf x}) = -4\pi n({\bf x}).
\end{equation}
For a spherical density $n({\bf x}) = n(r)$, this becomes
\begin{equation}
{1\over r^2}(r^2 V_H')' = V_H''(r) + {2\over r}V_H'(r) = -4\pi n(r),
\label{poisson:eq}
\end{equation}
where $n(r)$ is the radial particle (number) density,
normalized such that
\begin{equation}
N = \int n({\bf x}) \dd[3]{x} = \int_0^\infty 4\pi n(r) r^2 \dd r,
\label{poisson:norm}
\end{equation}
where $N$ is the number of electrons.

\subsubsection{Initial conditions}

Substituting (\ref{poisson:eq}) into (\ref{poisson:norm}) and integrating, we obtain
\begin{equation}
\lim_{r\to\infty} r^2 V_H'(r) = -N,
\end{equation}
from which it follows that the asymptotic behavior of $V_H'(r)$ is
\begin{equation}
V_H'(r) \sim -\frac{N}{r^{2}}, \qquad r \to \infty.
\label{poisson:V_H'}
\end{equation}
Integrating (\ref{poisson:V_H'}) and requiring $V_H \to 0$ as $r \to \infty$ then gives
the corresponding asymptotic behavior for $V_H(r)$,
\begin{equation} V_H(r) \sim \frac{N}{r}, \qquad r \to \infty.
\end{equation} For small $r$, the asymptotic behavior can be obtained by
expanding $n(r)$ about $r=0$: $n(r) = \sum_{j=0}^\infty {c_j r^j}$. Substituting into
Poisson equation \eqref{poisson:eq} gives
\begin{equation} (r^2 V_H')' = -4\pi \sum_{j=0}^\infty {c_j r^{j+2}}.
\end{equation} Integrating and requiring $V_H(0)$ to be finite then gives
\begin{equation}
\label{eq:dVsmallr}
V_H'(r) = -4\pi \sum_{j=0}^\infty { c_j \frac{r^{j+1}}{j+3} },
\end{equation}
with linear leading term, so that we have
\begin{equation}
\label{eq:dVsmallr2}
V_H'(r) \sim r, \qquad r \to 0.
\end{equation}
Integrating \eqref{eq:dVsmallr} then gives
\begin{equation}
\label{eq:Vsmallr}
V_H(r) = -4\pi \sum_{j=0}^\infty { c_j \frac{r^{j+2}}{(j+2)(j+3)} } + C,
\end{equation}
with leading constant term $C = V_H(0)$ determined by Coulomb's law:
\begin{equation}
\label{VH0}
V_H(0) = 4 \pi \int_0^\infty {r n(r) \dd r}.
\end{equation}
Finally, from \eqref{eq:dVsmallr2} we have that
\begin{equation}
\label{dVH0}
V_H'(0) = 0.
\end{equation}
So $V_{H}\to0$ as $r\to\infty$ and $V_{H}'(0)=0$. This asymptotic behavior provides the
initial values and derivatives for numerical integration in both inward and
outward directions.

\subsection{Kohn-Sham equations}\label{kseq}
The Kohn--Sham equations consist of the radial Schrödinger or Dirac equations
with an effective potential $V(r) = V_\inrm(r)$ given by (see, e.g.,~\cite{martinElectronicStructureBasic2004})
\begin{equation}
V_\inrm = V_H + V_{xc} + v,
\end{equation}
where $V_H$ is the Hartree potential given by the solution of the radial
Poisson equation (\ref{poisson:eq}), $V_{xc}$ is the exchange-correlation
potential, and $v = -\frac{Z}{r}$ is the nuclear potential.

The total energy is given by
\begin{equation}
E[n] = T_s[n] + E_H[n] + E_{xc}[n] + V[n],
\end{equation}
the sum of kinetic energy
\begin{equation}
T_s[n] = \sum_{nl} f_{nl} \varepsilon_{nl} -4\pi\int V_\inrm(r) n(r) r^2 \dd r,
\end{equation}
where $\varepsilon_{nl}$ are the Kohn-Sham eigenvalues,
Hartree energy
\begin{equation}
E_H[n] = 2\pi\int V_H(r) n(r) r^2 \dd r,
\end{equation}
exchange-correlation energy
\begin{equation}
E_{xc}[n] = 4\pi\int \varepsilon_{xc}(r; n) n(r) r^2 \dd r,
\end{equation}
where $\varepsilon_{xc}(r; n)$ is the exchange and correlation energy density,
and Coulomb energy
\begin{equation}
V[n] = 4\pi\int v(r) n(r) r^2 \dd r
    = -4\pi Z\int n(r) r \dd r
\end{equation}
with electronic density in the nonrelativistic case given by
\begin{equation}
    \label{eq-schr-density}
    n(r) = {1\over4\pi} \sum_{nl} f_{nl} {P_{nl}^2(r)\over r^2},
\end{equation}
where $P_{nl}$ is the radial wavefunction in \eqref{schroed_eq} and
$f_{nl}$ the associated electronic occupation. In the relativistic case, the
electronic density is given by
\begin{equation}
    \label{eq-dirac-density}
    n(r) = {1\over4\pi} \sum_{nls} f_{nls}
    {P_{nls}^2(r) + Q_{nls}^2(r) \over r^2},
\end{equation}
where $P_{nls}$ and $Q_{nls}$ are the two components of the Dirac
solution (\eqref{dirac_eq1},~\eqref{dirac_eq2})
and $f_{nls}$ is the occupation.
In both the above cases, $n(r)$ is the electronic particle density [electrons/volume], everywhere positive, as distinct from the electronic charge density $\rho(r)$ [charge/volume]: $\rho(r) = -n(r)$ in atomic units.

We adopt a self-consistent approach \cite{martinElectronicStructureBasic2004} to
solve for the electronic structure. Starting from an initial density $n_\inrm$
and corresponding potential $V_\inrm$, we solve the Schrödinger or Dirac
equation to determine the wavefunctions $R_{nl}$ or spinor components $P$ and
$Q$, respectively. From these, we construct a new density $n_\outrm$ and
potential $V_\outrm$. Subsequently, we update the input density and potential,
using for example a weighting parameter $\alpha\in[0, 1]$:
\begin{align}
n_\inrm \rightarrow \alpha n_\inrm + (1-\alpha) n_\outrm, \\
V_\inrm \rightarrow \alpha V_\inrm + (1-\alpha) V_\outrm.
\end{align}

\noindent This process is repeated until the difference of $V_\inrm$ and $V_\outrm$ and/or $n_\inrm$ and $n_\outrm$ is within a specified tolerance, at which point \textit{self-consistency} is achieved. This fixed-point iteration is known as the \textit{self-consistent field} (SCF) iteration. We employ an adaptive linear mixing scheme, with optimized weights for each component of the potential to construct new input potentials for successive SCF iterations.
To accelerate the convergence of the SCF iterations, under relaxed fixed
point iteration methods are used. In particular, the
code supports both \texttt{linear} \cite{novakAdaptiveAndersonMixing2023} and (default) Periodic \texttt{Pulay}
mixing schemes \cite{banerjeePeriodicPulayMethod2016}, which suitably accelerate convergence.
 In order to reduce the number of SCF iterations, we use a Thomas--Fermi (TF) approximation~\cite{oulneVariationSeriesApproach2011} for the initial density and potential:

\begin{subequations}\label{eq:tf-solution}
\begin{gather}
V(r) = -\frac{Z_\text{eff}(r)}{r}, \\
Z_\text{eff}(r) = Z\left(1 + \alpha\sqrt{x} + \beta x e^{-\gamma\sqrt{x}}\right)^2 e^{-2\alpha\sqrt{x}}, \\
x = r \left(\frac{128 Z}{9\pi^2}\right)^{1/3}, \\
\alpha = 0.7280642371, \quad \beta = -0.5430794693, \quad \gamma = 0.3612163121.
\end{gather}
\end{subequations}
The corresponding charge density is then
\begin{equation}
\label{tf:rho}
\rho(r) = -{1 \over 3 \pi^2} \left(-2 V(r)\right)^{3\over2}.
\end{equation}

We demonstrate the methodology with the local density approximation (LDA) and relativistic local-density approximation (RLDA) exchange and correlation functionals.
We note that the modular nature of the code and interface mechanism make it straightforward to incorporate functionals from other packages such as the library of exchange correlation functions, \texttt{libxc} \cite{marquesLibxcLibraryExchange2012,lehtolaRecentDevelopmentsLibxc2018}.
The parameters used in \texttt{featom} are taken from the same NIST benchmark data \cite{cohen1986AdjustmentFundamental1987} as the current state-of-the-art \texttt{dftatom} program for an accurate comparison. For the local density approximation,
\begin{equation}
V_{xc}(r;n) = {\dd \over \dd n}\left(n\varepsilon_{xc}^{LD}(n)\right),
\end{equation}
where the exchange and correlation energy density $\varepsilon_{xc}^{LD}$
can be written as~\cite{martinElectronicStructureBasic2004}
\begin{equation}
\varepsilon_{xc}^{LD}(n)=\varepsilon_x^{LD}(n)+\varepsilon_c^{LD}(n),
\end{equation}
with electron gas exchange term~\cite{martinElectronicStructureBasic2004}
\begin{equation}
\varepsilon_x^{LD}(n)=-{3\over4\pi}(3\pi^2 n)^{1\over3}
\end{equation}
and Vosko-Wilk-Nusair (VWN)~\cite{voskoAccurateSpindependentElectron1980} correlation term
\begin{multline}
\varepsilon_c^{LD}(n)\sim {A\over2}\left\{
\log\left(y^2\over Y(y)\right)
+{2b\over Q}\arctan\left(Q\over 2y+b\right) \right. \\ \left.
-{by_0\over Y(y_0)}\left[
   \log\left((y-y_0)^2\over Y(y)\right)
   +{2(b+2y_0)\over Q}\arctan\left(Q\over 2y+b\right)
   \right]
\right\},
\end{multline}
in which $y=\sqrt{r_s}$, $Y(y)=y^2+by+c$, $Q=\sqrt{4c-b^2}$, $y_0=-0.10498$,
$b=3.72744$, $c=12.9352$, $A=0.0621814$, and
\begin{equation}
r_s=\left(3\over4\pi n\right)^{1\over3}
\end{equation}
is the Wigner-Seitz radius, which gives the mean distance between electrons.
In the relativistic (RLDA) case, a correction to the LDA
exchange energy density and potential is given
by MacDonald and Vosko~\cite{macdonaldRelativisticDensityFunctional1979}:
\begin{subequations}
\begin{gather}
\varepsilon_x^{RLD}(n) = \varepsilon_x^{LD}(n) R, \\
R = 1-{3\over2}\left(\beta\mu-\log(\beta+\mu)\over\beta^2\right)^2, \notag \\
V_{x}^{RLD}(n) = V_{x}^{LD}(n) S, \\
S = {3\log(\beta+\mu)\over 2 \beta \mu} - \half, \notag
\end{gather}
\end{subequations}
where $\mu=\sqrt{1+\beta^2}$ and $\beta={(3\pi^2n)^{1\over3}\over c}
=-{4\pi \varepsilon_x^{LD}(n)\over 3c}$.

\section{Solution methodology}\label{methods}

Having described the Schrödinger, Dirac, and Kohn-Sham electronic structure
equations to be solved, and key solution properties, we now detail our
approach to solutions in a high-order finite-element basis.

\subsection{Radial Schrödinger equation}\label{ferschrod}

To recast \eqref{radial} in a manner that will facilitate our finite element
solution methodology, we make the substitution $P_{nl}(r) = rR_{nl}(r)$ to
obtain the canonical radial Schrödinger equation in terms of $P_{nl}(r)$ is
\begin{equation}
\label{schroed_eq}
-\frac{1}{2} P_{nl}''(r) + \left(V(r) + \frac{l(l+1)}{2r^2}\right)P_{nl}(r)
    = E P_{nl}(r)\,.
\end{equation}

\noindent  The corresponding normalization of $P(r)$ is

\begin{equation}
  \label{normPr}
\int_0^\infty P_{nl}^2(r) \dd r = 1.
\end{equation}

\subsubsection{Asymptotics}

The known asymptotic
behavior of $P_{nl}$ as $r\to0$ is~\cite{johnsonAtomicStructureTheory2007}

\begin{equation}
\label{sch_asympt_outward}
P_{nl}(r) \sim r^{l+1},
\end{equation}

\noindent where $P_{nl}$, being a solution of a homogeneous system of equations,
is only unique up to an arbitrary multiplicative constant. We
use~\eqref{schroed_eq} as our starting point but from now on drop the $nl$
index from $P_{nl}$ for simplicity:

\begin{equation}
  \label{eqsimplePnl}
-\frac{1}{2} P''(r) + \left(V(r) + \frac{l(l+1)}{2r^2}\right)P(r) = E P(r).
\end{equation}

In order to facilitate rapid convergence and application of desired boundary
conditions in a finite element basis, we generalize \eqref{schroed_eq} to solve
for $\tilde P=\frac{P(r)}{r^\alpha}$ for any chosen real exponent $\alpha\ge0$ by
substituting $P=r^{\alpha}\tilde P$ to obtain

\begin{equation}
\label{schroed_alpha}
-\frac{1}{2} {1\over r^{2\alpha}} \left(r^{2\alpha}\tilde P'(r)\right)'
+ \left(V(r) + \frac{l(l+1) - \alpha(\alpha-1)}{2r^2}\right) \tilde P(r)
= E \tilde P(r).
\end{equation}

\noindent We note that:

\begin{itemize}

\item For $\alpha=0$ we get $\tilde P(r) = \frac{P(r)}{r^0}=P(r)$ and~\eqref{schroed_alpha} reduces to~\eqref{schroed_eq}.

\item For $\alpha=1$ we get $\tilde P(r) = \frac{P(r)}{r^1}=R(r)$ and~\eqref{schroed_alpha} reduces to \eqref{radial}.

\item Finally, for $\alpha=l+1$, (which corresponds to the known asymptotic \eqref{sch_asympt_outward}) we obtain $\tilde P(r) = \frac{P(r)}{r^{l+1}}$, which tends to a nonzero value at the origin for all $l$ and~\eqref{schroed_alpha}
becomes

\begin{equation}
  \label{alphalpone}
-\frac{1}{2} {1\over r^{2(l+1)}} \left(r^{2(l+1)}\tilde P'(r)\right)'
+ V(r) \tilde P(r)
= E \tilde P(r).
\end{equation}

\end{itemize}

\subsubsection{Weak form}

To obtain the weak form, we multiply both sides of \eqref{schroed_alpha} by a
test function $v(r)$ and integrate from $0$ to $\infty$. In addition, to facilitate
the construction of a symmetric bilinear form, we multipy by a factor $r^{2\alpha}$
to get

\begin{flalign}
\int_0^\infty \left[ -\frac{1}{2} \left(r^{2\alpha}\tilde P'(r)\right)' v(r)
+ \left(V(r) + \frac{l(l+1) - \alpha(\alpha-1)}{2r^2}\right) \tilde P(r) v(r)
r^{2\alpha}
    \right] \dd r  \\\nonumber
    = E \int_0^\infty \tilde P(r) v(r) r^{2\alpha} \dd r.
\end{flalign}

\noindent We can now integrate by parts to obtain

\begin{align}
\int_0^\infty \left[\frac{1}{2} r^{2\alpha} \tilde P'(r) v'(r)
+ \left(V(r) + \frac{l(l+1) - \alpha(\alpha-1)}{2r^2}\right) \tilde P(r) v(r)
r^{2\alpha}
    \right] \dd r  \\\nonumber
    -\half\left[r^{2\alpha}\tilde P'(r) v(r)\right]_0^\infty
    = E \int_0^\infty \tilde P(r) v(r) r^{2\alpha} \dd r.
\end{align}

Setting the boundary term to zero,

\begin{equation}
    \label{boundary_term}
\left[r^{2\alpha}\tilde P'(r) v(r)\right]_0^\infty = 0,
\end{equation}

we then obtain the desired symmetric weak formulation

\begin{align}
\label{schroed_weak_form}
\int_0^\infty \left[\half \tilde P'(r) v'(r)
+ \left(V(r) + \frac{l(l+1) - \alpha(\alpha-1)}{2r^2}\right) \tilde P(r) v(r)
    \right] r^{2\alpha} \dd r\\\nonumber
    = E \int_0^\infty \tilde P(r) v(r) r^{2\alpha} \dd r.
\end{align}

\noindent As discussed below, by virtue of our choice of $\alpha$ and boundary conditions on
$v(r)$, the vanishing of the boundary term \eqref{boundary_term} imposes no
natural boundary conditions on $\tilde P$.
\subsubsection{Discretization}

To discretize (\ref{schroed_weak_form}), we introduce finite element basis functions $\phi_i(r)$ to form trial and test functions

\begin{align}\label{eqn:P1_expand}
\tilde P(r) &= \sum_{j=1}^{N} c_j \phi_j(r) \qquad \mathrm{and}\qquad v(r)    = \phi_i(r),
\end{align}

\noindent and substitute into \eqref{schroed_weak_form}. In so doing, we obtain a generalized
eigenvalue problem

\begin{equation}
\sum_{j=1}^N H_{ij} c_j = E \sum_{j=1}^N S_{ij} c_j,
\end{equation}

\noindent where $H_{ij}$ is the Hamiltonian matrix,

\begin{equation}
\label{eqn:sym_bc2}
H_{ij} = \int_0^\infty \left[\half \phi_i'(r) \phi_j'(r)
+ \phi_i(r) \left(V + \frac{l(l+1) - \alpha(\alpha-1)}{2r^2}\right)
    \phi_j(r) \right] r^{2\alpha} \dd r,
\end{equation}
and $S_{ij}$ is the overlap matrix,
\begin{equation}
S_{ij} = \int_0^\infty \phi_i(r) \phi_j(r) r^{2\alpha} \dd r.
\end{equation}

\subsubsection{Basis}
While, by virtue of the weak formulation, the above discretization can employ any basis in Sobolev space $H^1$ satisfying the required boundary conditions (including standard $C^0$ FE bases, $C^1$ Hermite, and $C^k$ B-splines), we employ a high-order $C^0$ spectral element (SE) basis [34,35] in the present work. FE bases consist of local piecewise polynomials. SE bases are a form of FE basis constructed to enable well-conditioned, high polynomial-order discretization via definition over Lobatto nodes rather than uniformly spaced nodes within each finite element (mesh interval). In the present work, we employ polynomial orders up to $p = 31$.

In the electronic structure context, SE bases have a number of desirable properties, including:
\begin{enumerate}
\item Polynomial completeness and associated systematic convergence with increasing number and/or polynomial order of basis functions.
\item Well conditioned to high polynomial order by virtue of definition over Lobatto nodes.
\item Exponential convergence with respect to polynomial order, enabling high accuracy with a small basis.
\item Applicable to general nonsingular as well as singular Coulombic potentials, whether fixed or self-consistent.
\item Applicable to bound as well as excited states.
\item Can use uniform as well as nonuniform meshes.
\item Basis function evaluation, differentiation, and integration are inexpensive, enabling fast and accurate matrix element computations.
\item Dirichlet boundary conditions are readily imposed by virtue of cardinality, i.e.,
$$\phi_i(x_j) = \delta_{ij}$$
where $x_j$ is a nodal point of the basis.
\item Derivative boundary conditions are readily imposed via the weak formulation.
\item Diagonal overlap matrix using Lobatto quadrature, thus enabling solution of a standard rather than generalized eigenvalue problem, reducing both storage requirements and time to solution.
\end{enumerate}

Properties (1)-(4), (7), and (10) are advantageous relative to Slater-type and Gaussian-type bases, which can be nontrivial to converge; can become ill-conditioned as the number of basis functions is increased, limiting accuracies attainable in practice; and produce a non-diagonal overlap matrix, requiring solution of a generalized rather than standard matrix eigenvalue problem.

Properties (2) and (10) are advantageous relative to B-spline bases, which are generally employed at lower polynomial orders in practice and give rise to a non-diagonal overlap matrix, thus requiring solution of a generalized rather than standard matrix eigenvalue problem.

\subsubsection{Boundary conditions}\label{bcrschrod}
Since we seek bound states, which vanish as $r\to\infty$, we impose a
homogeneous
Dirichlet boundary condition on $v(r)$ and $\tilde P(r)$ at $r=\infty$ by employing a
finite element basis $\{\phi_{i}\}$ satisfying this condition.

For $\alpha=0$, $\tilde P(r)=P(r)=0$ at $r=0$ for all quantum numbers $l$ according
to \eqref{sch_asympt_outward}. Thus we impose a homogeneous Dirichlet boundary
condition on $v(r)$ and $\tilde P(r)$ at $r=0$ by employing a finite element
basis $\{\phi_{i}\}$ satisfying this condition, whereupon the boundary term
\eqref{boundary_term} as a whole vanishes, consistent with the weak formulation
\eqref{schroed_weak_form}.

For $\alpha=1$, $\tilde P(r)= \frac{P(r)}{r}=R(r)=0$ at $r=0$ for all quantum numbers
$\ell>0$ according to \eqref{sch_asympt_outward}. However, $\tilde P(r)\neq0$ at $r=0$
for $\ell=0$. Thus a homogeneous Dirichlet boundary condition cannot be imposed for
$\ell=0$. However, for $\alpha>0$, the $r^{2\alpha}$ factor in \eqref{boundary_term} vanishes
at $r=0$, whereupon the boundary term as a whole vanishes, consistent with the
weak formulation \eqref{schroed_weak_form}, regardless of boundary condition
on $v(r)$ and $\tilde P(r)$ at $r=0$. Hence for $\alpha > 0$, we impose a
homogeneous Dirichlet boundary condition on $v(r)$ and $\tilde P(r)$
at $r=\infty$ only, by employing a finite element basis
$\{\phi_{i}\}$ satisfying this condition. This is sufficient due to the singularity of
the associated Sturm-Liouville problem.

Similarly, for $\alpha=\ell+1$, $\tilde P(r)=\frac{P(r)}{r^{\ell+1}}\neq0$ at $r=0$ for all
quantum numbers $\ell$ according to \eqref{sch_asympt_outward} and, since $\alpha>0$, we
again impose a homogeneous Dirichlet boundary condition on $v(r)$ and
$\tilde P(r)$ at $r=\infty$ only.

In practice, the \texttt{featom} code defaults to $\alpha=0$ with homogeneous Dirichlet
boundary conditions at $r=0$ and $r=\infty$.

Note that, while $\alpha=0$ yields a rapidly convergent formulation in the context of
the Schrödinger equation, it does not suffice in the context of the Dirac
equation, as discussed in Section~\ref{bcrdirac}.

\subsection{Radial Dirac equation}\label{ferdirac}
For small $r$, the central potential has the form $V(r) = -Z/r + Z_1 + O(r)$,
which gives rise to the following asymptotic behavior at the origin for
$Z \ne 0$
(Coulombic)~\cite{grantDiracOperatorFinite2008,zabloudilElectronScatteringSolid2006}:
\begin{subequations}
\label{dirac_asympt_outward_singu}
\begin{flalign}
&P_{n\kappa}(r) \sim r^{\beta}, \\
&Q_{n\kappa}(r) \sim r^{\beta} {c(\beta+\kappa)\over Z}, \\
&\beta = \sqrt{\kappa^2-\left(Z\over c\right)^2}.
\end{flalign}
\end{subequations}
For $Z=0$ (nonsingular) the asymptotic is, for $\kappa < 0$~\cite{grantDiracOperatorFinite2008}
\begin{subequations}
\label{dirac_asympt_outward_nosing_lezero}
\begin{align}
\label{dirac_asympt_outward_nonsing1}
P_{n\kappa}(r) &\sim  r^{l+1}, \\
\label{dirac_asympt_outward_nonsing1b}
Q_{n\kappa}(r) &\sim  r^{l+2} {E + Z_1 \over c(2l+3)},
\end{align}
\end{subequations}
and for $\kappa > 0$
\begin{subequations}
\label{dirac_asympt_outward_nosing_gezero}
\begin{align}
\label{dirac_asympt_outward_nonsing2}
P_{n\kappa}(r) &\sim -r^{l+2} {E + Z_1 \over c(2l+1)}, \\
\label{dirac_asympt_outward_nonsing2b}
Q_{n\kappa}(r) &\sim  r^{l+1}.
\end{align}
\end{subequations}
For large $r$, assuming $V(r) \to 0$ as $r\to\infty$, the asymptotic
is~\cite{johnsonAtomicStructureTheory2007}
\begin{subequations}
\begin{flalign}
    &P_{n\kappa}(r) \sim e^{-\lambda r},\\
    &Q_{n\kappa}(r) \sim - \sqrt{-{E\over E + 2c^2}} P_{n\kappa}(r),\\
&\lambda = \sqrt{c^2 -{(E+c^2)^2\over c^2}} =\sqrt{-2E - {E^2\over c^2}}.
\end{flalign}
\end{subequations}
Consistent with the coupled equations \eqref{dirac_eq1} and \eqref{dirac_eq2},
the Dirac Hamiltonian can be written as (see,
e.g.,~\cite[(7)]{fischerBsplineGalerkinMethod2009})
\begin{equation}
\label{H_orig}
    H = \begin{pmatrix}
        V(r) & c\left(-{\dv{r}}+{\kappa\over r}\right) \\
        c\left({\dv{r}}+{\kappa\over r}\right) & V(r) - 2c^2 \\
        \end{pmatrix}.
\end{equation}
The corresponding eigenvalue problem is then
\begin{equation}
    \label{PQ_eig}
    H
        \begin{pmatrix}
        P(r) \\
        Q(r) \\
        \end{pmatrix}
        =
    E
        \begin{pmatrix}
        P(r) \\
        Q(r) \\
        \end{pmatrix}.
\end{equation}
To discretize, we expand the solution vector in a basis:
\begin{equation}
   \label{PQ_expand}
   \begin{pmatrix}
   P(r) \\
   Q(r)
   \end{pmatrix}
       =
   \sum_{j=1}^{2N} c_j
       \begin{pmatrix}
       \phi_j^a(r) \\
       \phi_j^b(r)
       \end{pmatrix},
\end{equation}

\noindent where $a$ and $b$ denote the upper and lower components of the vector.
We have $2N$ basis functions (degrees of freedom), $N$ for each wave function
component. As such, the function $P(r)$ is expanded in terms of basis functions
$\phi_i^a(r)$ and the function $Q(r)$ in terms of $\phi_i^b(r)$. However, these two
expansions are not independent but are rather  coupled via coefficients $c_i$.

\subsubsection{Squared Hamiltonian formulation}\label{sqrdirac}

The above eigenvalue formulation of the radial Dirac equation can be solved
using the finite element method. However, due to the fact that the operator is
not bounded from below, one obtains spurious eigenvalues in the spectrum
\cite{tupitsynSpuriousStatesDirac2008}. To eliminate spurious states, we work
with the square of the Hamiltonian
\cite{wallmeierUseSquaredDirac1981,kutzelniggBasisSetExpansion1984}, which is
bounded from below, rather than the Hamiltonian itself. Since the eigenfunctions
of the square of an operator are the same as those of the operator itself, and
the eigenvalues are the squares, the approach is straightforward and enables
direct and efficient solution by the finite element method.

Let us derive the
equations for the squared radial Dirac Hamiltonian. First we shift the energy by
$c^2$ to obtain the relativistic energy, making the Hamiltonian more symmetric,
using \eqref{H_orig} to get
\begin{equation}
\label{H_orig2}
    H + \mathbb{I}c^2 = \begin{pmatrix}
        V(r) + c^2 & c\left(-{\dv{r}}+{\kappa\over r}\right) \\
        c\left({\dv{r}}+{\kappa\over r}\right) & V(r) - c^2 \\
        \end{pmatrix}.
\end{equation}
Then we square the Hamiltonian using \eqref{PQ_eig} to obtain the following
equations to solve for $P(r)$ and $Q(r)$:
\begin{equation}
    \label{PQ_eig2}
    (H+\mathbb{I}c^2)^2
        \begin{pmatrix}
        P(r) \\
        Q(r) \\
        \end{pmatrix}
        =
    (E+c^2)^2
        \begin{pmatrix}
        P(r) \\
        Q(r) \\
        \end{pmatrix},
\end{equation}
\noindent where $\mathbb{I}$ is the $2\times2$ identity matrix.

\noindent As can be seen, the eigenvectors $P(r)$ and $Q(r)$ are the same as before but
the eigenvalues are now equal to $(E+c^2)^2$ and the original non-relativistic
energies $E$ can be obtained by taking the square root of these new eigenvalues
and subtracting $c^2$.

\subsubsection{Weak form}

We follow a similar approach as for the Schrödinger equation: instead of solving
for $P(r)$ and $Q(r)$, we solve for $\tilde P(r) = {P(r)\over r^\alpha}$ and
$\tilde Q(r) = {Q(r)\over r^\alpha}$, which introduces a parameter $\alpha$ that can be
chosen to facilitate rapid convergence and the application of the desired
boundary conditions in a finite element basis. We then multiply the eigenvectors
by $r^\alpha$ to obtain $P(r)$ and $Q(r)$.

Now we can write the finite element formulation as

\begin{subequations}    \label{squared_fem1}
  \begin{flalign}
      A &=
    \int_0^\infty\!\!
\begin{pmatrix} \phi_i^a(r) & \phi_i^b(r) \\ \end{pmatrix}
    r^\alpha (H+\mathbb{I}c^2)^2 r^\alpha
        \begin{pmatrix}
        \phi_j^a(r) \\
        \phi_j^b(r)
        \end{pmatrix}
        \dd r, \\
      S &=
    \int_0^\infty
\begin{pmatrix} \phi_i^a(r) & \phi_i^b(r) \\ \end{pmatrix}
    r^\alpha r^\alpha
        \begin{pmatrix}
        \phi_j^a(r) \\
        \phi_j^b(r)
        \end{pmatrix}
        \dd r, \\
      A x &= (E+c^2)^2 S x.
\end{flalign}
\end{subequations}
This is a generalized eigenvalue problem, with eigenvectors $x$
(coefficients of $(\tilde P(r),\tilde Q(r))$), eigenvalues $(E+c^2)^2$,
and $2 \times 2$ block matrices $A$ and $S$.
Note that the basis functions from both sides were multiplied by $r^\alpha$
due to the substitutions $P(r) = r^\alpha \tilde P(r)$ and $Q(r) = r^\alpha
\tilde Q(r)$. We can denote the middle factor in the integral for $A$ as $H' = r^{\alpha} \left(H+\mathbb{I}c^2\right)^2 r^{\alpha}$
and compute it using \eqref{H_orig2} with rearrangement of $r^\alpha$ factors as
follows:
\begin{subequations}\label{eq:SqasymH}
\begin{align}
    H' &= r^{\alpha} \left(H+\mathbb{I}c^2\right)^2 r^{\alpha} \\
       &= r^{2\alpha} \begin{pmatrix}
        V(r) + c^2 & c \left(-{\dv{r}}+{\kappa-\alpha\over r}\right) \\
        c \left({\dv{r}}+{\kappa+\alpha\over r}\right) & V(r) - c^2 \\
        \end{pmatrix}^2.
\end{align}
\end{subequations}

\noindent Let
\begin{equation}
H'= \begin{pmatrix}
    H^{11} & H^{12} \\
    H^{21} & H^{22}
    \end{pmatrix}.
\end{equation}

\noindent Then

\begin{subequations}
\begin{align}
    H^{11} &= r^{2\alpha}\left(V(r) + c^2\right)^2 + r^{2\alpha}c^2\left(-{\dv[2]{r}}-{2\alpha\over r} {\dv{r}}+ \Phi\right), \\
    H^{12} &= r^{2\alpha}c\left(2{\left(\kappa-\alpha\right)\over r} V(r) - 2V(r) {\dv{r}} - V'(r) \right), \\
    H^{21} &= r^{2\alpha}c\left(2{\left(\kappa+\alpha\right)\over r} V(r) + 2V(r) {\dv{r}} + V'(r) \right), \\
    H^{22} &= r^{2\alpha}\left(V(r) - c^2\right)^2 + r^{2\alpha}c^2 \left(-{\dv[2]{r}}-{2\alpha\over r} {\dv{r}}+ \Phi\right),
\end{align}
\end{subequations}

\noindent where $\displaystyle\Phi = \frac{\left(\kappa\left(\kappa+1\right)-\alpha\left(\alpha-1\right)\right)}{r^2}$.

\noindent The full derivation of these expressions is given in \ref{App:SqHamDeriv}.

\noindent The usual approach when applying the finite element method to a system of equations
is to choose the basis functions in the following form:
\begin{subequations}\label{eq:basisFunc}
\begin{align}
\phi_i^a(r)  &= \begin{cases}
                \pi_i(r) & \text{for $i=1, \dots, N$}, \\
                0        & \text{for $i=N+1, \dots, 2N$}.
            \end{cases} \\
\phi_i^b(r)  &= \begin{cases}
                0            & \text{for $i=1, \dots, N$}, \\
                \pi_{i-N}(r) & \text{for $i=N+1, \dots, 2N$}.
            \end{cases}
\end{align}
\end{subequations}

\noindent Substituting (\ref{eq:basisFunc}) into (\ref{squared_fem1}), we obtain the following expressions after simplification (\ref{App:weakSqDH}):
\begin{subequations}
\begin{flalign}
A &= \begin{pmatrix}
    A^{11} & A^{12} \\
    A^{21} & A^{22}
    \end{pmatrix}, \\
S &= \begin{pmatrix}
    S^{11} & 0 \\
    0 & S^{22} \\
    \end{pmatrix},
\end{flalign}
\end{subequations}
\noindent with components given by
\begin{subequations}
\begin{align}
  \label{eq:dirbc}
A^{11}_{ij} &= \int_0^\infty \left( c^2 \pi_i'(r) \pi_j'(r) + \left(\left(V(r) + c^2\right)^2 + c^2 \Phi\right) \pi_i(r) \pi_j(r) \right) r^{2\alpha} \textrm{d} r, \\
A^{22}_{ij} &= \int_0^\infty \left( c^2 \pi_i'(r) \pi_j'(r) + \left(\left(V(r) - c^2\right)^2 + c^2 \Phi\right) \pi_i(r) \pi_j(r) \right) r^{2\alpha} \textrm{d} r, \\
A^{12}_{ij} &= \int_0^\infty c V(r) \left(+\pi_i'(r)\pi_j(r) - \pi_i(r) \pi_j'(r) + 2{\kappa\over r} \pi_i(r) \pi_j(r) \right) r^{2\alpha} \textrm{d} r,\\
A^{21}_{ij} &= \int_0^\infty c V(r) \left(-\pi_i'(r)\pi_j(r) + \pi_i(r) \pi_j'(r) + 2{\kappa\over r} \pi_i(r) \pi_j(r) \right) r^{2\alpha} \textrm{d} r,\\
S^{11}_{ij} &= S^{22}_{ij} = \int_0^\infty \pi_i(r) \pi_j(r) r^{2\alpha} \textrm{d} r,
\end{align}
\end{subequations}

\noindent where $\displaystyle\Phi = \frac{\kappa\left(\kappa+1\right)-\alpha\left(\alpha-1\right)}{r^2}$.

\noindent These are the expressions implemented in the \texttt{featom} code.
\subsubsection{Basis}\label{basis}
Because we solve the squared Dirac Hamiltonian problem \eqref{PQ_eig2}, whose
spectrum has a lower bound, rather than the Dirac Hamiltonian problem
\eqref{PQ_eig}, whose spectrum is unbounded above and below, we can employ a
high-order $C^0$ SE basis
\cite{pateraSpectralElementMethod1984,hafeezReviewApplicationsSpectral2023}, as
in the Schrödinger case, as discussed in Section~\ref{bcrschrod}, with exponential
convergence to high accuracy and no spurious states.

As discussed in Section~\ref{introduction}, a number of approaches have been
developed over the past few decades to solve \eqref{PQ_eig} directly without spurious
states, with varying degrees of success. Perhaps most common is to use different
bases for large and small components, $P(r)$ and $Q(r)$, of the Dirac
wavefunction, e.g., \cite{dyallKineticBalanceVariational1990,shabaevDualKineticBalance2004,beloyApplicationDualkineticbalanceSets2008,fischerBsplineGalerkinMethod2009,sunComparisonRestrictedUnrestricted2011,jiaoDevelopmentKineticallyAtomically2021}. And among these approaches, perhaps most
common is to impose some form of ``kinetic balance,'' e.g., \cite{sunComparisonRestrictedUnrestricted2011,grantRelativisticQuantumTheory2007} and
references therein, on the bases for $P$ and $Q$. However, imposing such
conditions significantly complicates the basis, increasing the cost of
evaluations and matrix element integrals, and while highly effective in practice
is not guaranteed to eliminate spurious states all cases \cite{sunComparisonRestrictedUnrestricted2011}. In any case, for
a $C^0$ basis as employed in the present work, such balance cannot be imposed
since it produces basis functions which are discontinuous and thus not in
Sobolev space $H^1$ as required (Section~\ref{bcrschrod}). Another approach which has proven
successful in the context of B-spline bases is to use different orders of
B-splines as bases for $P$ and $Q$ \cite{fischerBsplineGalerkinMethod2009,zatsarinnyDBSR_HFBsplineDirac2016,fischerBSplineAtomicStructure2021}.

However, this too creates additional complexity \cite{fischerBSplineAtomicStructure2021}, did not eliminate all spurious states in our implementation using $p$ and $p+1$ degree SE bases for $P$ and $Q$, respectively, nor eliminate all spurious states in previous work \cite{igarashiBSplineExpansionsRadial2006}. Thus we opt instead for the squared Hamiltonian approach in the present work in order to allow the robust and efficient use of a high-order $C^0$ SE basis, with all desired properties enumerated in Section~\ref{bcrschrod}. In addition, the squared Hamiltonian approach allows
direct solution for just the positive (electron) states, rather than having to
solve for both positive and negative, e.g., \cite{zatsarinnyDBSR_HFBsplineDirac2016}, thus saving computation.

Finally, as discussed in Section~\ref{bcrdirac}, we note that by solving for
$\tilde{P} = P/r^\alpha$ and $\tilde{Q} = Q/r^\alpha$, with $\alpha$ set according to the known
asymptotic as $r \rightarrow 0$, rather than for $P$ and $Q$ themselves, we obtain rapid
convergence for singular Coulomb as well as nonsingular potentials using our $C^0$
SE basis. On the other hand, singular Coulomb potentials have posed significant
difficulty for B-spline bases \cite{zatsarinnyDBSR_HFBsplineDirac2016}. However, since B-splines have polynomial
completeness to specified degree, like SE bases, solving for $\tilde{P}$ and
$\tilde{Q}$ rather than $P$ and $Q$ directly should yield rapid convergence for
such singular potentials using B-spline bases as well.

\subsubsection{Boundary conditions}\label{bcrdirac}

To construct a symmetric weak form, the following two boundary terms are set to
zero in the derivation (\ref{App:weakSqDH}):

\begin{subequations}
    \label{dirac_bterms}
\begin{align}
    c^2    r^{2\alpha} \pi_i(r) \pi_j'(r) \Biggr|^{\infty}_0 &= 0 \,, \\
    c V(r) r^{2\alpha} \pi_i(r) \pi_j(r)  \Biggr|^{\infty}_0 &= 0 \,.
\end{align}
\end{subequations}

As in the Schrödinger case, since we seek bound states which vanish as
$r\to\infty$, we impose a homogeneous Dirichlet boundary condition on the test
function $v(r)$ and solutions $\tilde P(r)$ and $\tilde Q(r)$ at $r=\infty$ by
employing a finite element basis $\{ \pi_{i} \}$ satisfying this condition.

Unlike the Schrödinger case, however, the appropriate exponent $\alpha$ and
boundary condition at $r=0$ depends on the regularity of the potential $V(r)$.

For nonsingular potentials $V(r)$, $P(r)\sim r^{\ell + 1}$ and $Q(r)\sim
r^{\ell+2}$ as $r\to0$ for relativistic quantum numbers $\kappa<0$ while
$P(r)\sim r^{\ell+2}$ and $Q(r)\sim r ^{\ell+1}$ for $\kappa>0$ according
to \eqref{dirac_asympt_outward_nosing_lezero} and
\eqref{dirac_asympt_outward_nosing_gezero}. Hence, as in the Schrödinger
case, for $\alpha=0$, $\tilde P(r)=P(r)=0$ and $\tilde Q(r)=Q(r)=0$ at $r=0$ for all
quantum numbers $\kappa$ and $\ell$. Thus we impose a homogeneous Dirichlet
bounday condition on $v(r)$, $\tilde P(r)$, and $\tilde Q(r)$ at $r=0$ by
employing a finite element basis $\{ \pi_{i} \}$ satisfying this condition, whereupon
the boundary terms \eqref{dirac_bterms} vanish, consistent with the weak
formulation \eqref{squared_fem1}. This is the default in the \texttt{featom}
code for such potentials.

For singular potentials $V(r)$ with leading term $-\frac{Z}{r}$ (Coulombic),
however, $P(r)$ and $Q(r)$ have leading terms varying as $r^{\beta}$ as $r\to0$
for all relativistic quantum numbers $\kappa$ according to
\eqref{dirac_asympt_outward_singu}. However, while $\beta>1$ for $|\kappa|>1$,
we have
$0.74<\beta<0.99998$ for $\kappa=\pm 1$ (for $1\leq Z \leq 92$), so that $P(r)$
and $Q(r)$ have non-polynomial behavior at small $r$ and divergent derivatives
at $r=0$, leading to numerical difficulties for methods attempting to compute
them directly. To address this issue, we leverage the generality of the
formulation \eqref{squared_fem1} to solve for $\tilde P(r) = P(r)/r^\alpha$ and
$\tilde Q(r) = Q(r)/r^\alpha$ with $\alpha = \beta$, rather than solving for
$P(r)$ and $Q(r)$ directly. With $\alpha=\beta$, $\tilde P(r)$ and $\tilde
Q(r)$ have polynomial behavior at small $r$ and bounded nonzero values at $r=0$
for all $\kappa$, including $\kappa=\pm1$, thus eliminating the aforementioned
numerical difficulties completely and facilitating rapid convergence in a
polynomial basis. Finally, for $\alpha = \beta$, the $r^{2\alpha}$ factors in
the boundary terms \eqref{dirac_bterms} vanish at $r=0$, whereupon the boundary
terms as a whole vanish, consistent with the weak formulation
\eqref{squared_fem1}, regardless of boundary condition on $v(r)$, $\tilde
P(r)$, and $\tilde Q(r)$ at $r=0$. Hence, for $\alpha=\beta$, we impose  a
homogeneous Dirichlet boundary condition at $r=\infty$ only, by employing a
finite element basis $\{ \pi_{i} \}$ satisfying this condition. This is the default
in the \texttt{featom} code for such potentials.

For integrals involving non-integer $\alpha$, we employ Gauss-Jacobi quadrature
for accuracy and efficiency, while for integrals involving only integer
exponents, we use Gauss-Legendre quadrature with the exception of overlap
integrals where we employ Gauss-Lobatto quadrature to obtain a diagonal overlap
matrix.

\section{Results and discussion}\label{results}
To demonstrate the accuracy and performance of the \texttt{featom}
implementation of the above described finite element formulation, we present
results for fixed potentials as well as self-consistent DFT calculations, with
comparisons to analytic results where available and to the state-of-the-art
\texttt{dftatom} solver otherwise. Code outputs are collected in
\ref{App:SysRes}.

\subsection{Coulombic systems}
The accuracy of the Schrödinger and Dirac solvers was verified using the
Coulomb potential $V = -\frac{Z}{r}$ for $Z=92$ (uranium). Eigenvalues are given by
the corresponding analytic formula \cite{certikDftatomRobustGeneral2013}. All
eigenvalues with $n < 7$ are used for the study.  The reference eigenvalues
here are from the analytic solutions: $E_{nl}=-\frac{Z^{2}}{2n^{2}}$ for the
Schrödinger equation, and

\begin{subequations}
  \begin{align}
    E_{n\kappa} &= \frac{c^{2}}{\sqrt{1+\frac{(Z/c)^{2}}{(n-|\kappa|+\beta)^{2}}}}-c^{2},\\
    \beta&=\sqrt{\kappa^{2}-(Z/c)^{2}}
  \end{align}
\end{subequations}
\noindent for the Dirac equation.

\begin{figure}
\centering
\subfloat[$p$ study for sum of eigenvalues]{
\label{fig:coulomb_schroed_conv}
\includegraphics[width=0.5\linewidth]{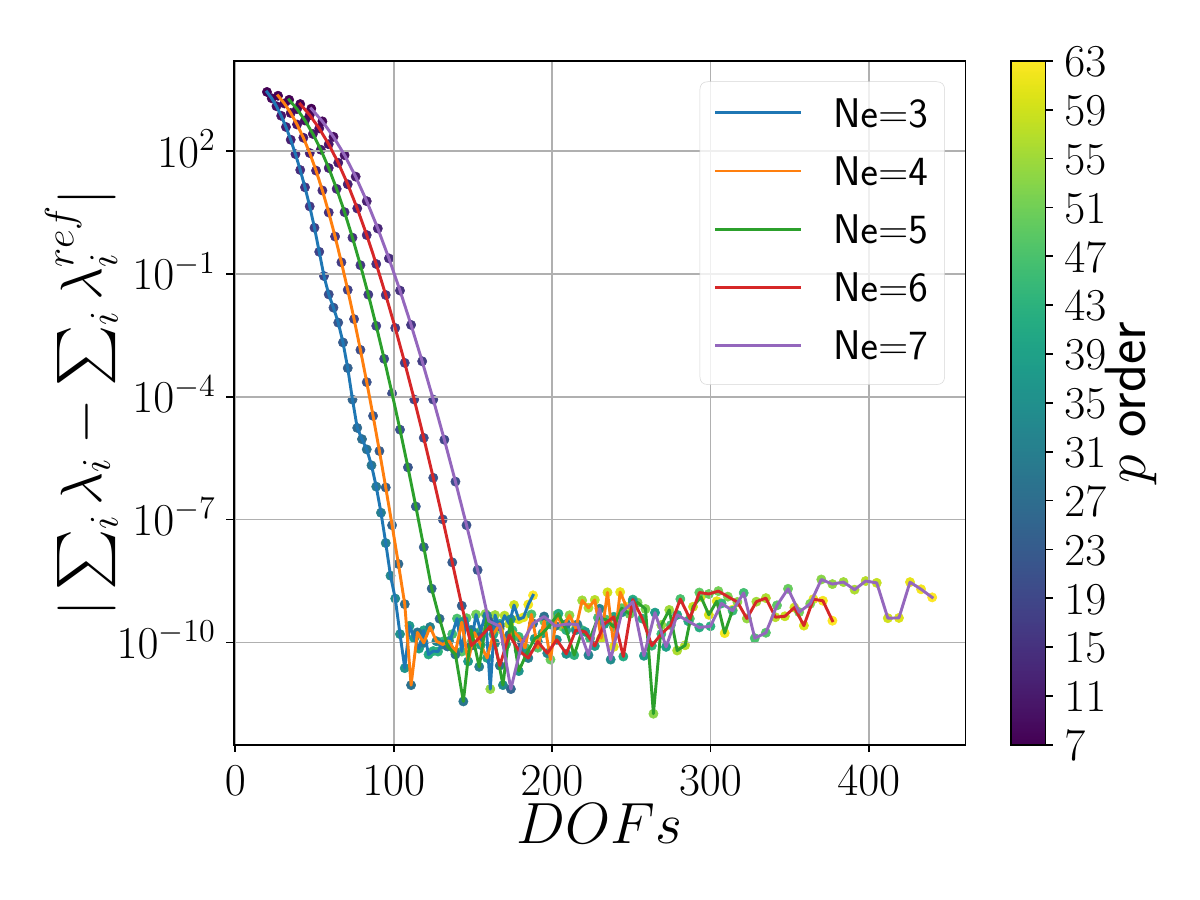}
}
\subfloat[$r_\mathrm{max}$ study for sum of eigenvalues]{
\label{fig:coulomb_schroed_rmax}
\includegraphics[width=0.5\linewidth]{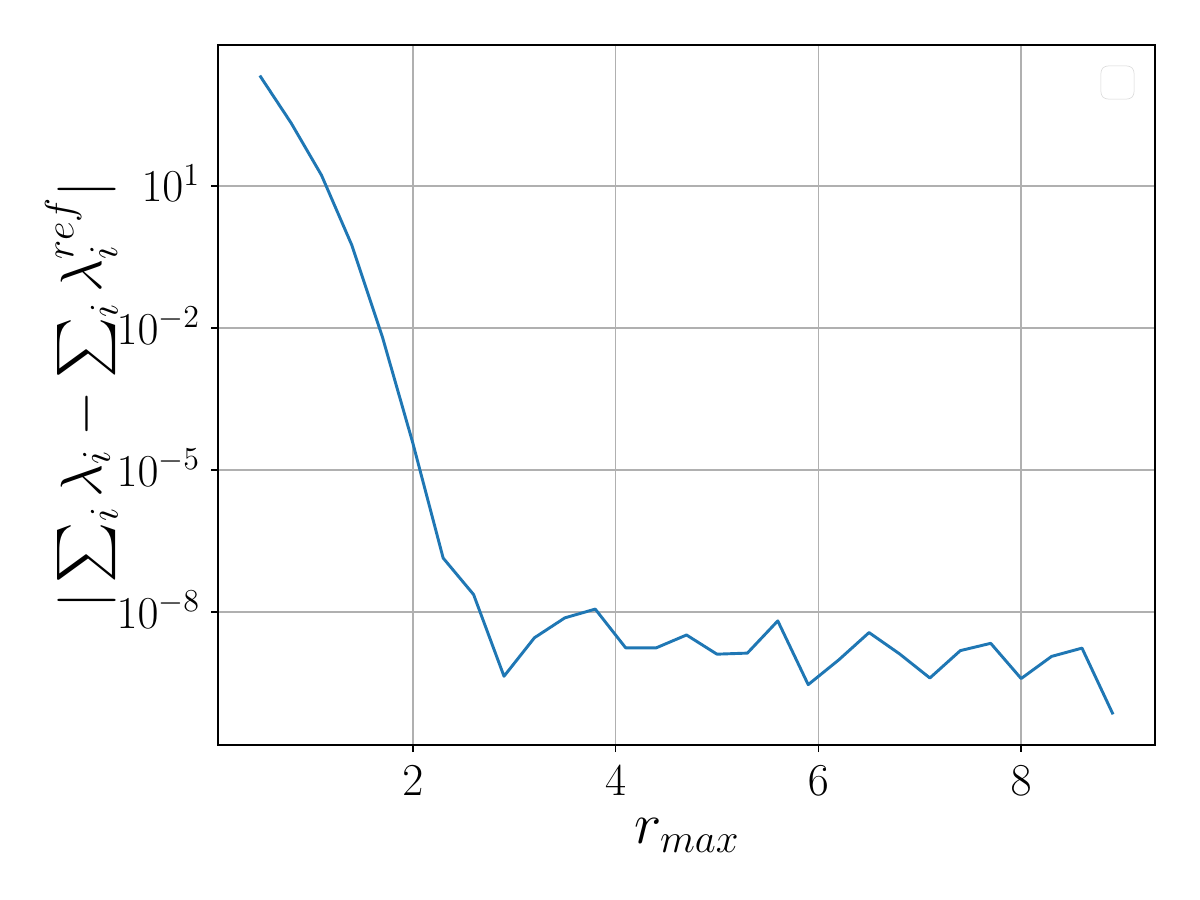}
}

\subfloat[$r_\mathrm{max}$ study for eigenvalues]{
\label{fig:coulomb_schroed_rmax_eig}
\includegraphics[width=0.5\linewidth]{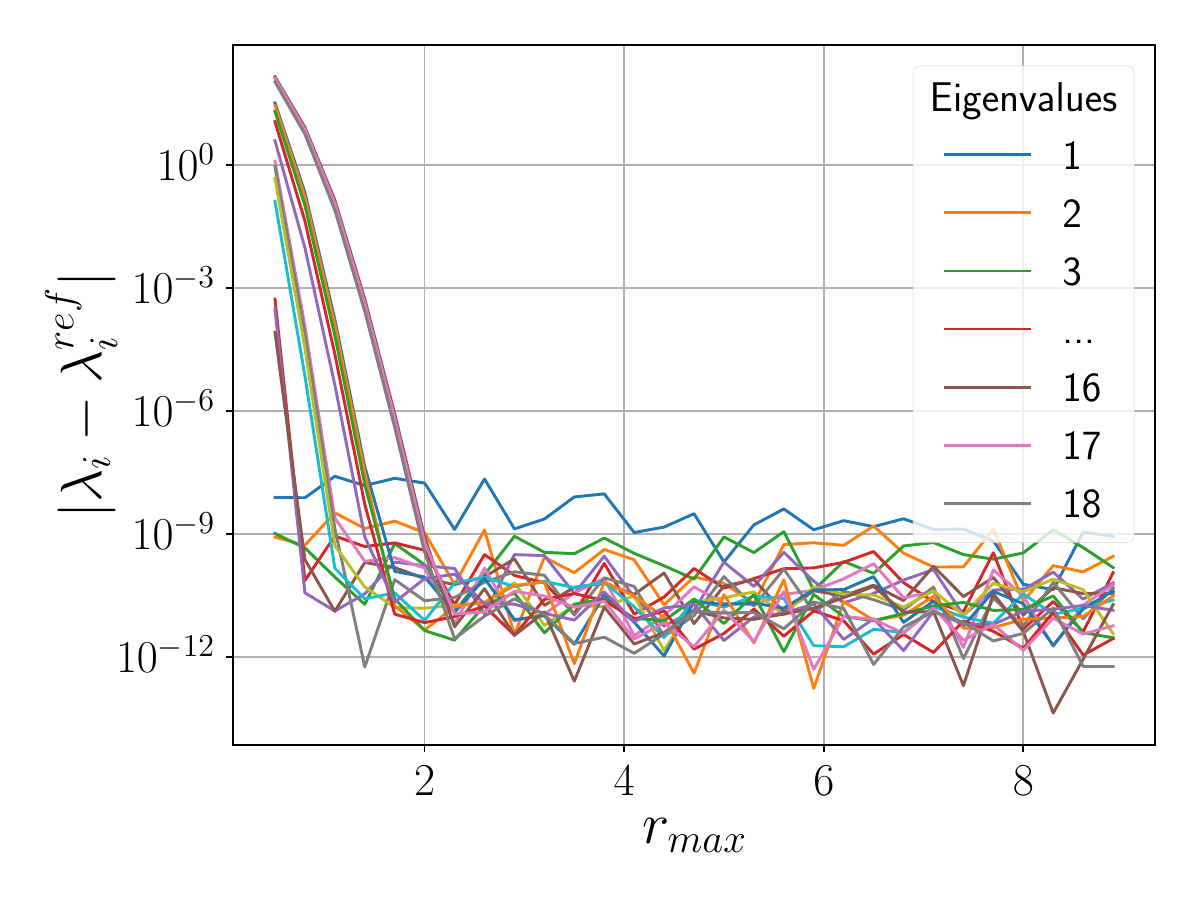}
}
\subfloat[$N_{e}$ study for sum of eigenvalues]{
\label{fig:coulomb_schroed_ne}
\includegraphics[width=0.5\linewidth]{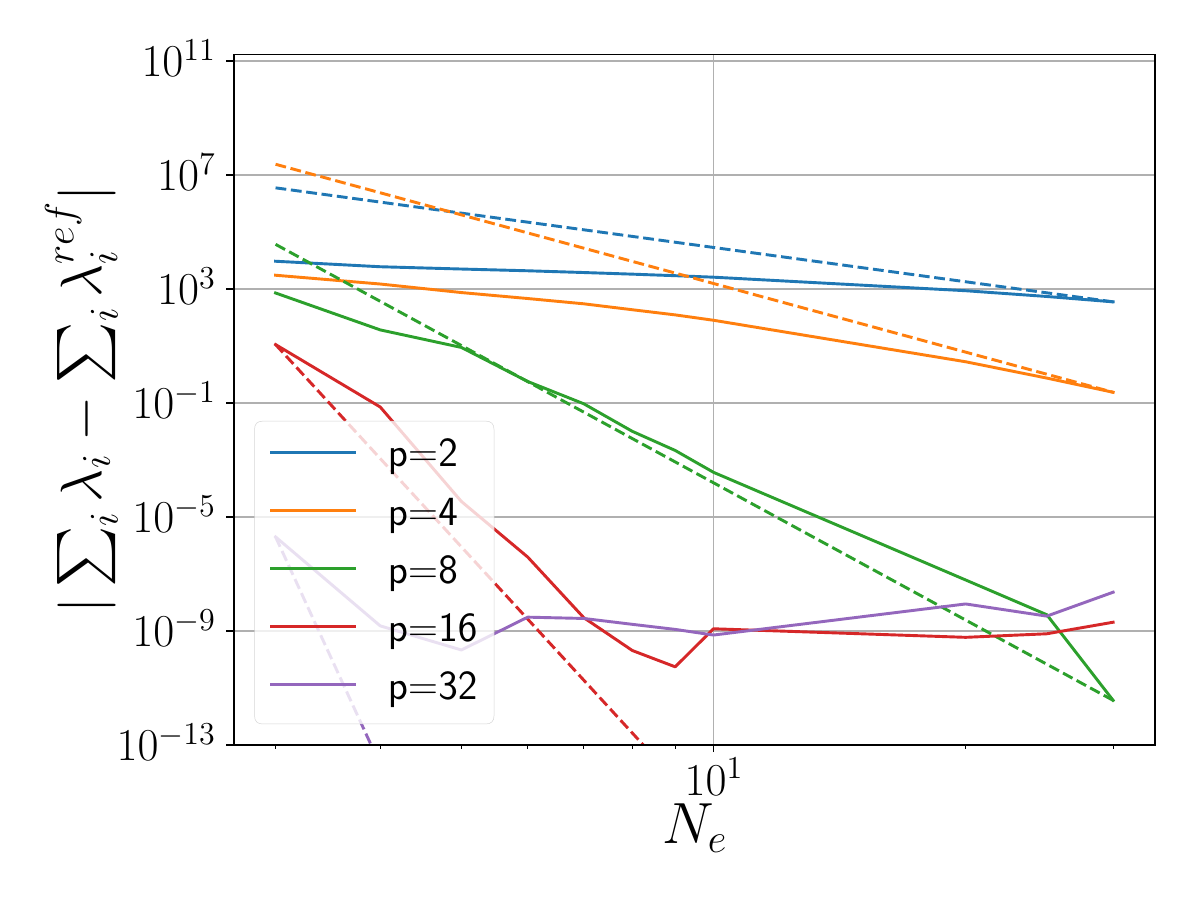}
}
\caption{Convergence studies for the Schrödinger equation with a Coulomb potential with $Z=92$.}
\end{figure}

Figure~\ref{fig:coulomb_schroed_conv} shows the convergence of the Schrödinger total energy
error with respect to the polynomial order $p$ for different numbers of elements
$N_{e}$. The graph, when observed for a given number of elements, forms a
straight line on the log-linear scale, showing that the error decreases
exponentially with the polynomial order $p$ until it reaches the
numerical precision limit of approximately $10^{-9}$.

For five elements, we considered the behavior of the error with respect to
$r_\mathrm{max}$. Figure~\ref{fig:coulomb_schroed_rmax} shows the
total energy error. It is observed that $r_\mathrm{max} \ge 10$ results in an
error of $10^{-8}$ or less. The eigenvalues converge to $10^{-9}$ for
$r_\mathrm{max} \ge 10$, as depicted in Figure~\ref{fig:coulomb_schroed_rmax_eig}.

The theoretical convergence rate for the finite element method is given by
$N_{e}^{-2p}$. Figure~\ref{fig:coulomb_schroed_ne} shows the error with respect to $N_{e}$, juxtaposed with the theoretical convergence represented as a
dotted line. The slope of the solid lines is used to determine the rate of
convergence. In all instances, we observe that the theoretical convergence rate
is achieved. For polynomial orders $p\le8$, it is attained asymptotically, while
for $p>8$, it approaches the limiting numerical precision of
around $10^{-9}$.
\begin{figure}
\centering
\subfloat[$p$ study for sum of eigenvalues]{
\label{fig:coulomb_dirac_conv}
\includegraphics[width=0.5\linewidth]{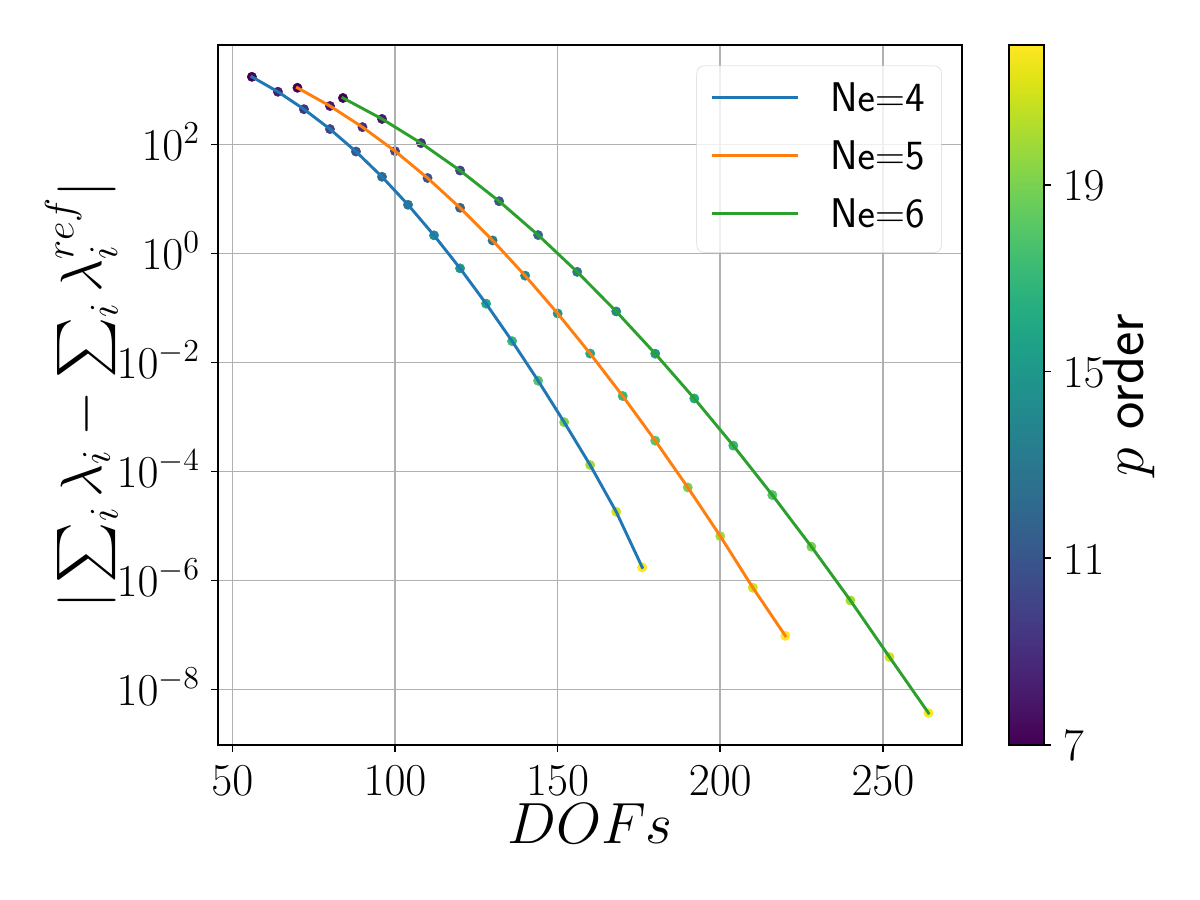}
}
\subfloat[$r_\mathrm{max}$ study for sum of eigenvalues]{
\label{fig:coulomb_dirac_rmax}
\includegraphics[width=0.5\linewidth]{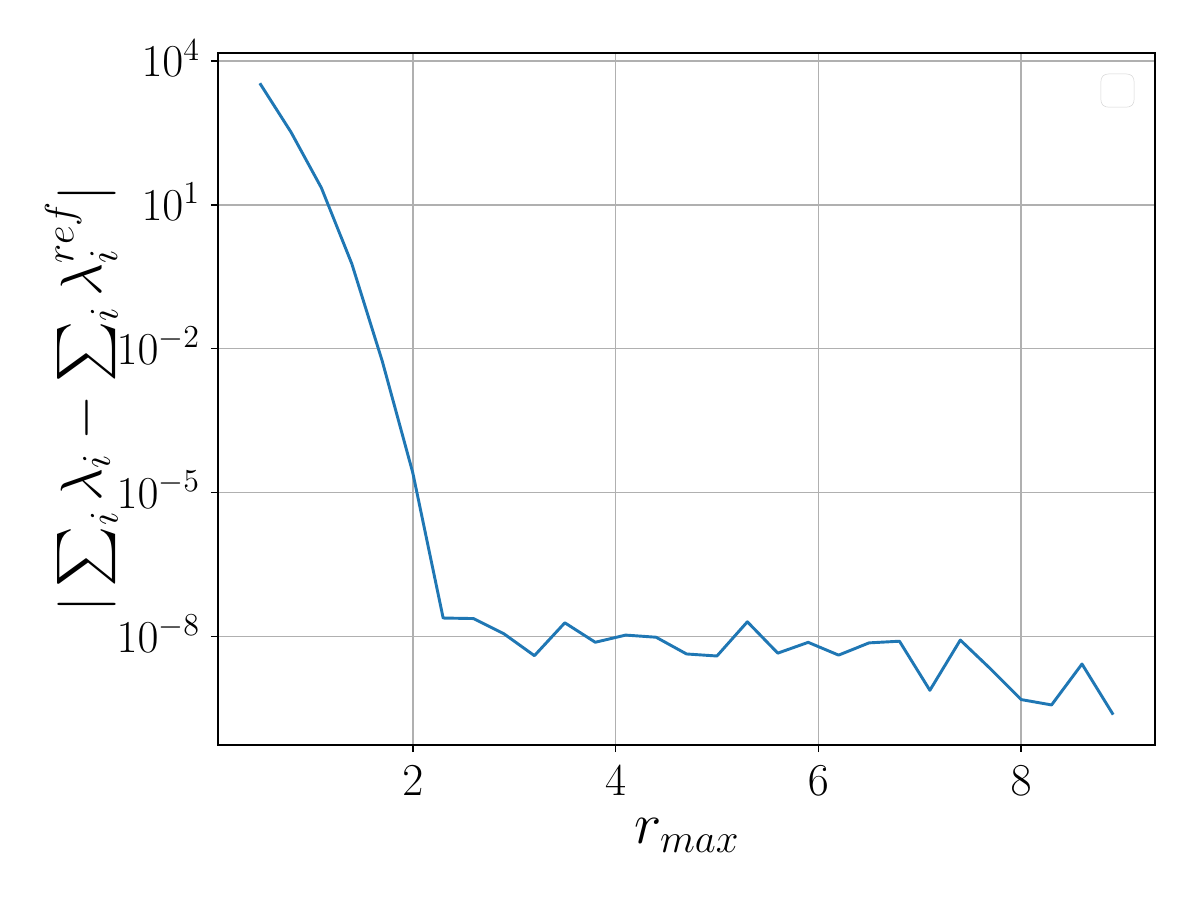}
}

\subfloat[$r_\mathrm{max}$ study for eigenvalues]{
\label{fig:coulomb_dirac_rmax_eig}
\includegraphics[width=0.5\linewidth]{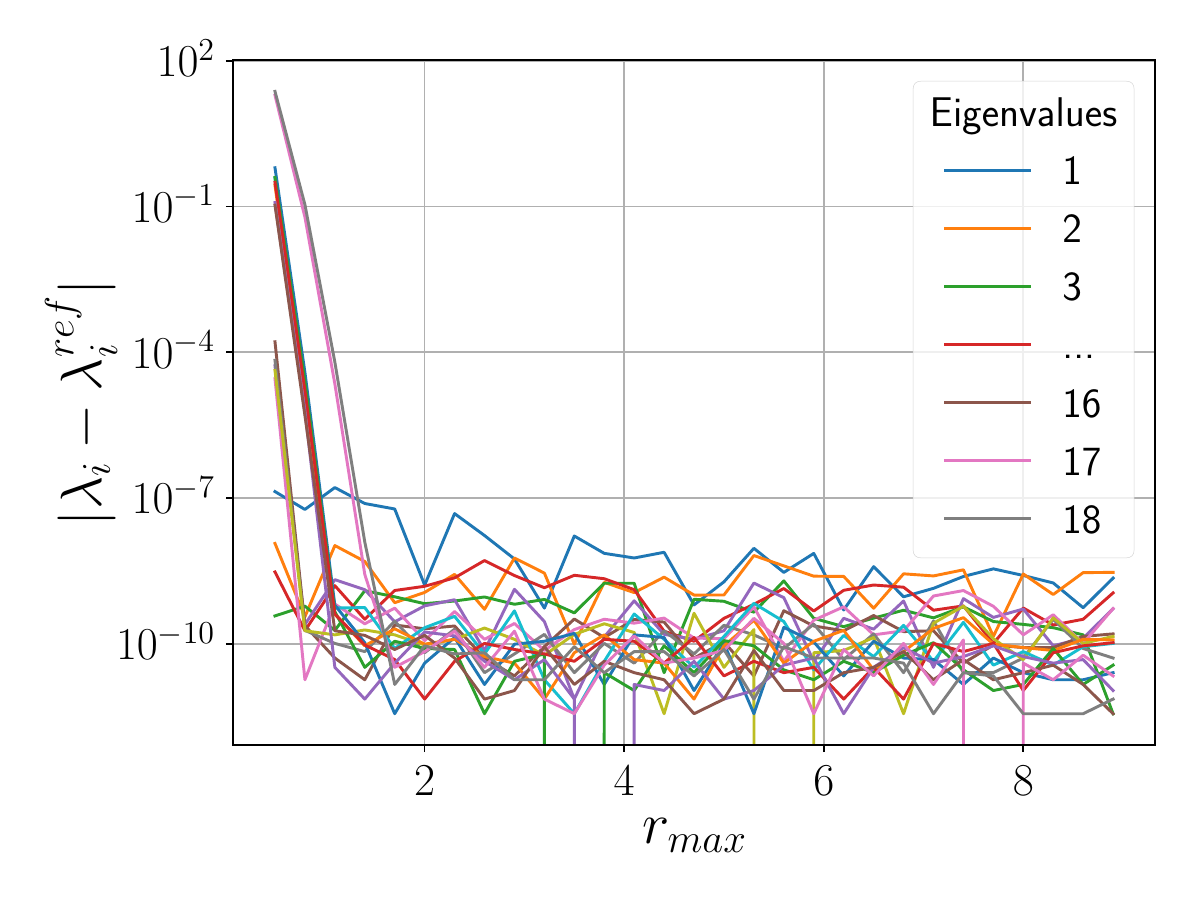}
}
\subfloat[$N_{e}$ study for sum of eigenvalues]{
\label{fig:coulomb_dirac_ne}
\includegraphics[width=0.5\linewidth]{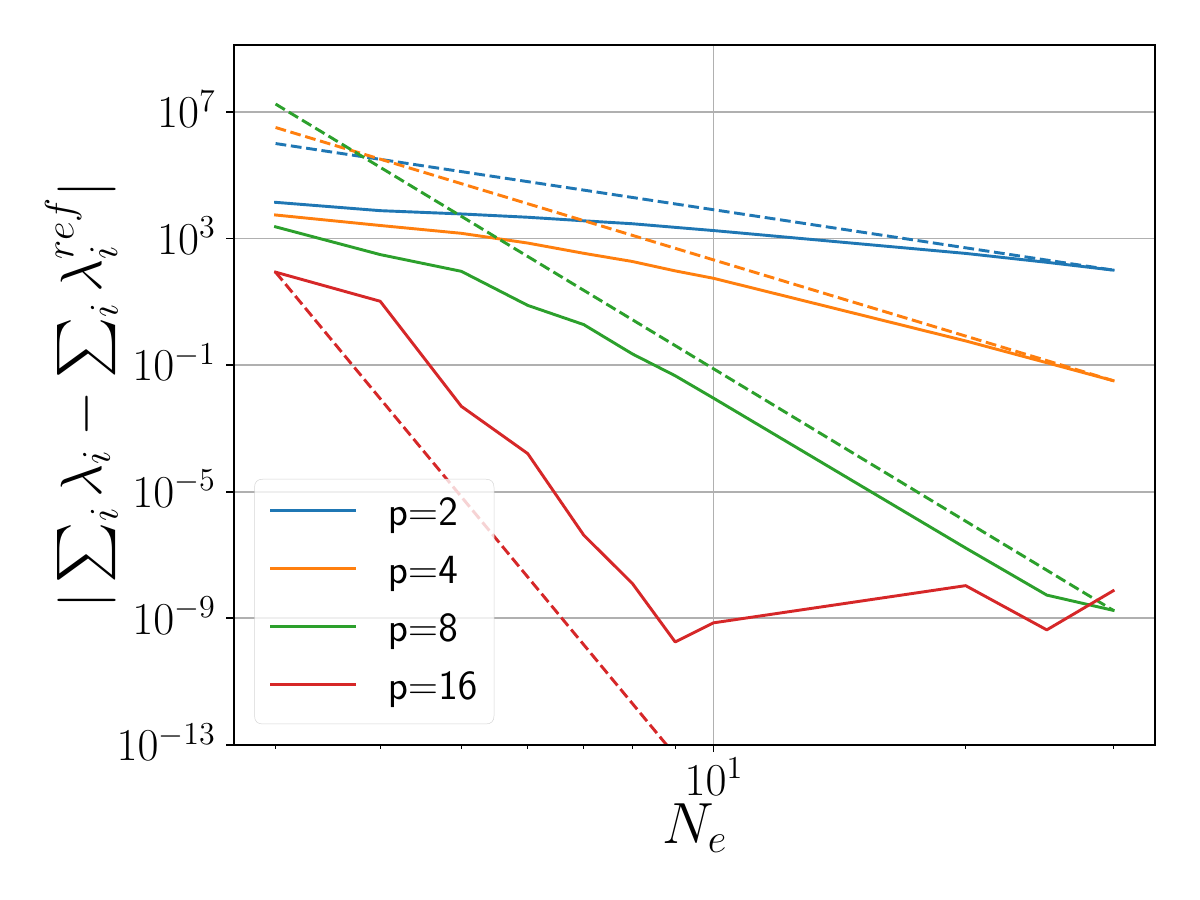}
}
\caption{Convergence studies for the Dirac equation with a Coulomb potential with $Z=92$.}
\end{figure}

Figure~\ref{fig:coulomb_dirac_conv} shows the error in
the Dirac total energy with respect to $p$ for various $N_{e}$. We find the same exponential
relationship between the error and polynomial order $p$.

As before, for five elements, we consider the error with respect to $r_\mathrm{max}$. Figure~\ref{fig:coulomb_dirac_rmax} shows the total Dirac energy error,
and it is observed that $r_\mathrm{max} \ge 10$ results in an error of $10^{-8}$
or less. The eigenvalues converge to $10^{-8}$ for $r_\mathrm{max} \ge 10$, as
depicted in Figure~\ref{fig:coulomb_dirac_rmax_eig}.

Figure~\ref{fig:coulomb_dirac_ne} shows the error as $N_{e}$ is increased
alongside the theoretical convergence represented as a dotted line. In all
instances, we observe that the theoretical convergence rate is achieved. For
polynomial orders $p\le8$, it is attained asymptotically, while for $p>8$, it approaches the limiting numerical precision of around $10^{-9}$.

\subsection{Quantum harmonic oscillator}

Next, we consider a nonsingular potential: the harmonic oscillator potential given by
$V(r)=\frac{1}{2}\omega^{2}r^{2}$. For the Schrödinger equation, we take $\omega=1$ and
compare against the exact values given by

\begin{align}
  E_{nl} = \omega(2n-l-\frac{1}{2}).
\end{align}

\begin{figure}
\centering
\subfloat[$p$ study for sum of eigenvalues]{
\label{fig:harmonic_schroed_conv}
\includegraphics[width=0.5\linewidth]{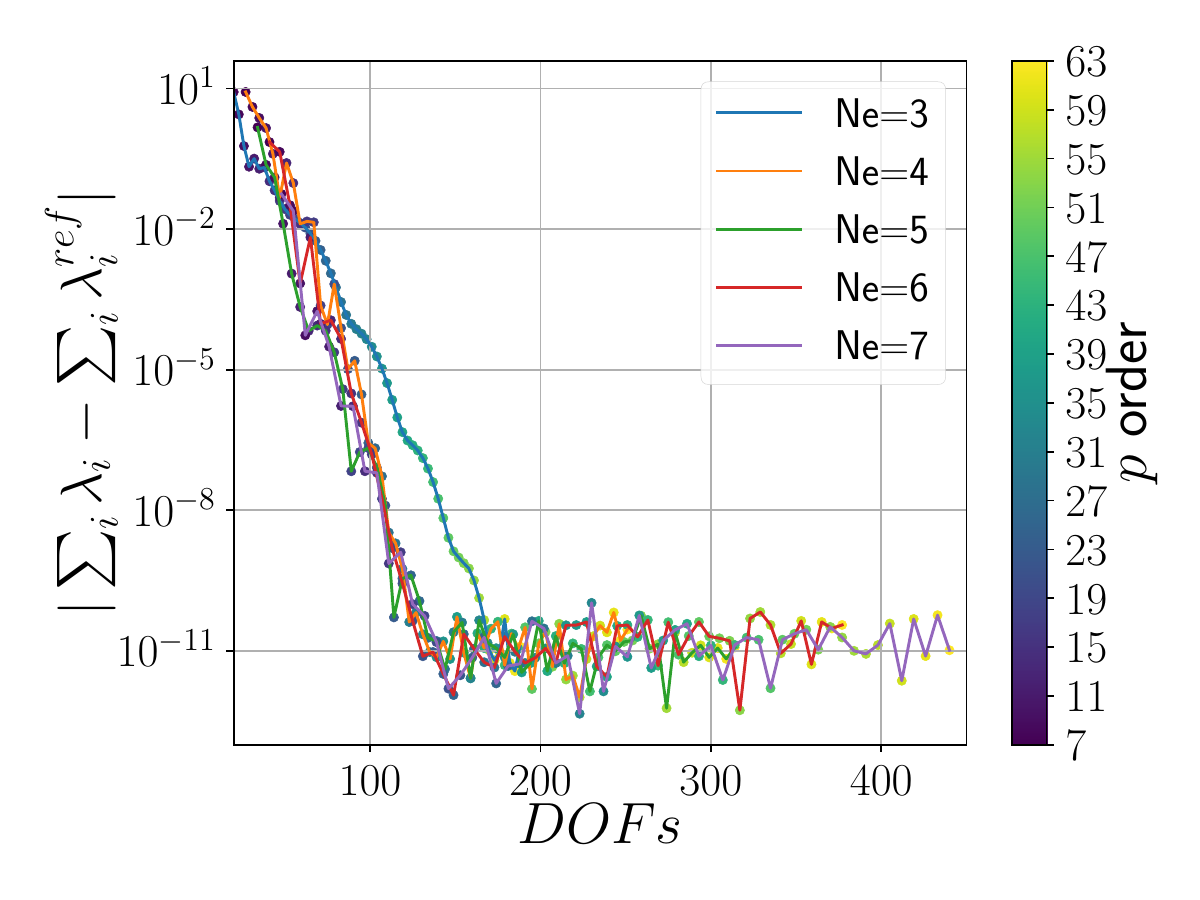}
}
\subfloat[$r_\mathrm{max}$ study for sum of eigenvalues]{
\label{fig:harmonic_schroed_rmax}
\includegraphics[width=0.5\linewidth]{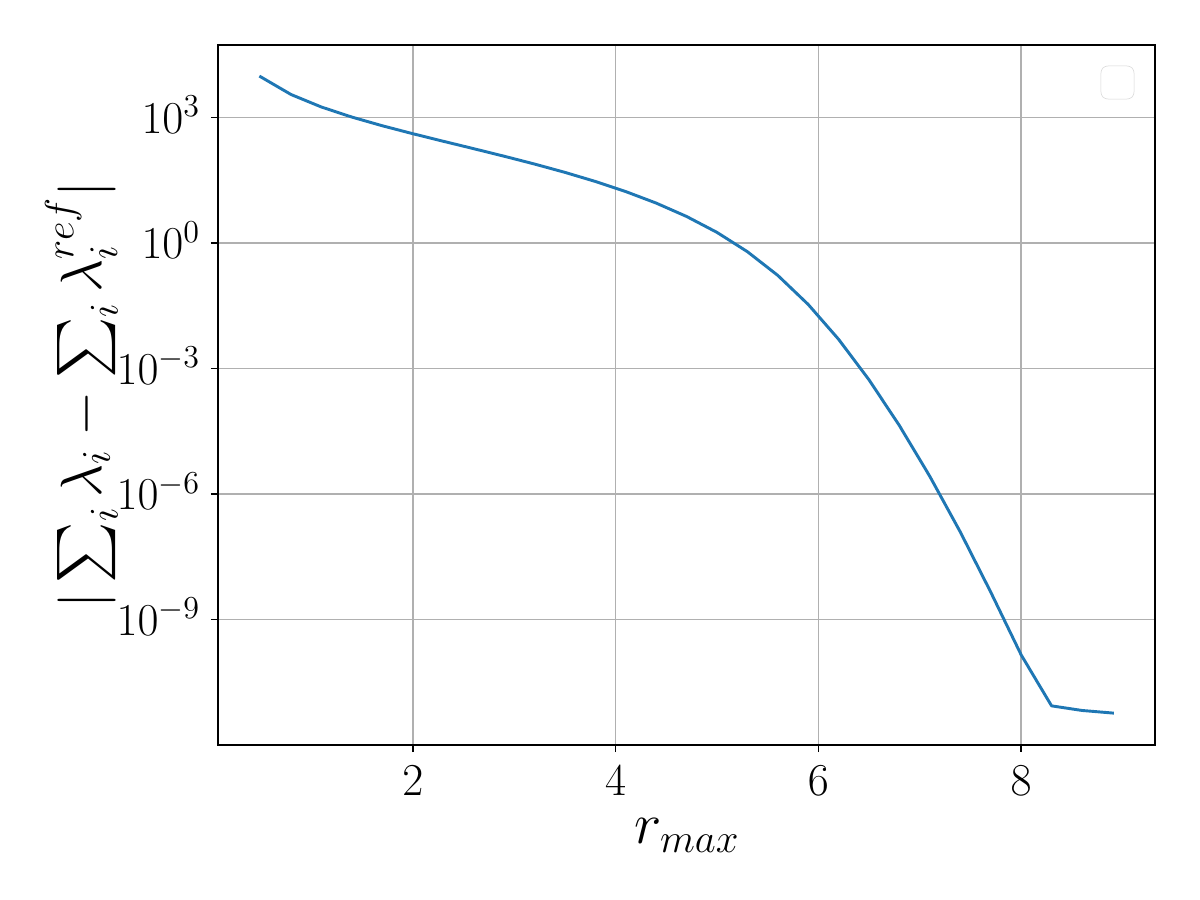}
}

\subfloat[$r_\mathrm{max}$ study for eigenvalues]{
\label{fig:harmonic_schroed_rmax_eig}
\includegraphics[width=0.5\linewidth]{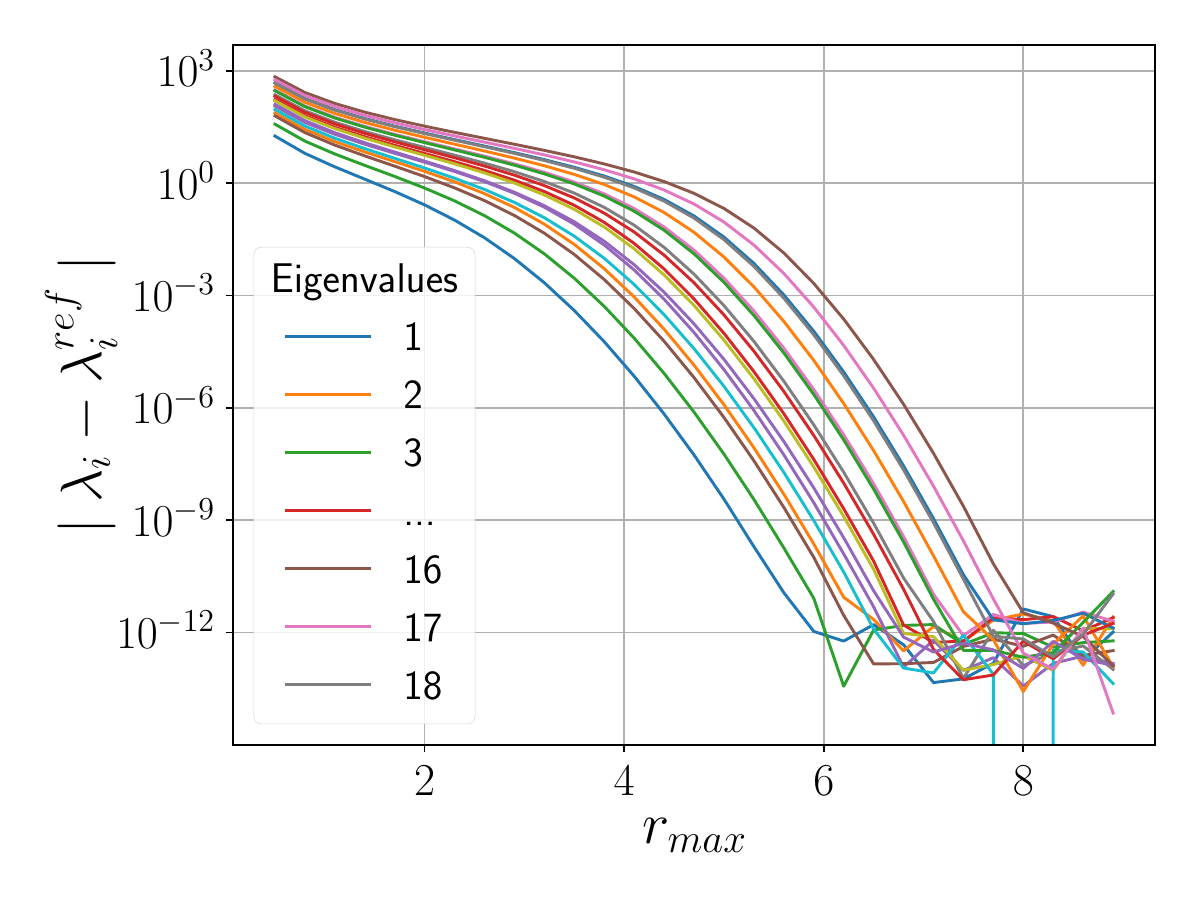}
}
\subfloat[$N_{e}$ study for sum of eigenvalues]{
\label{fig:harmonic_schroed_ne}
\includegraphics[width=0.5\linewidth]{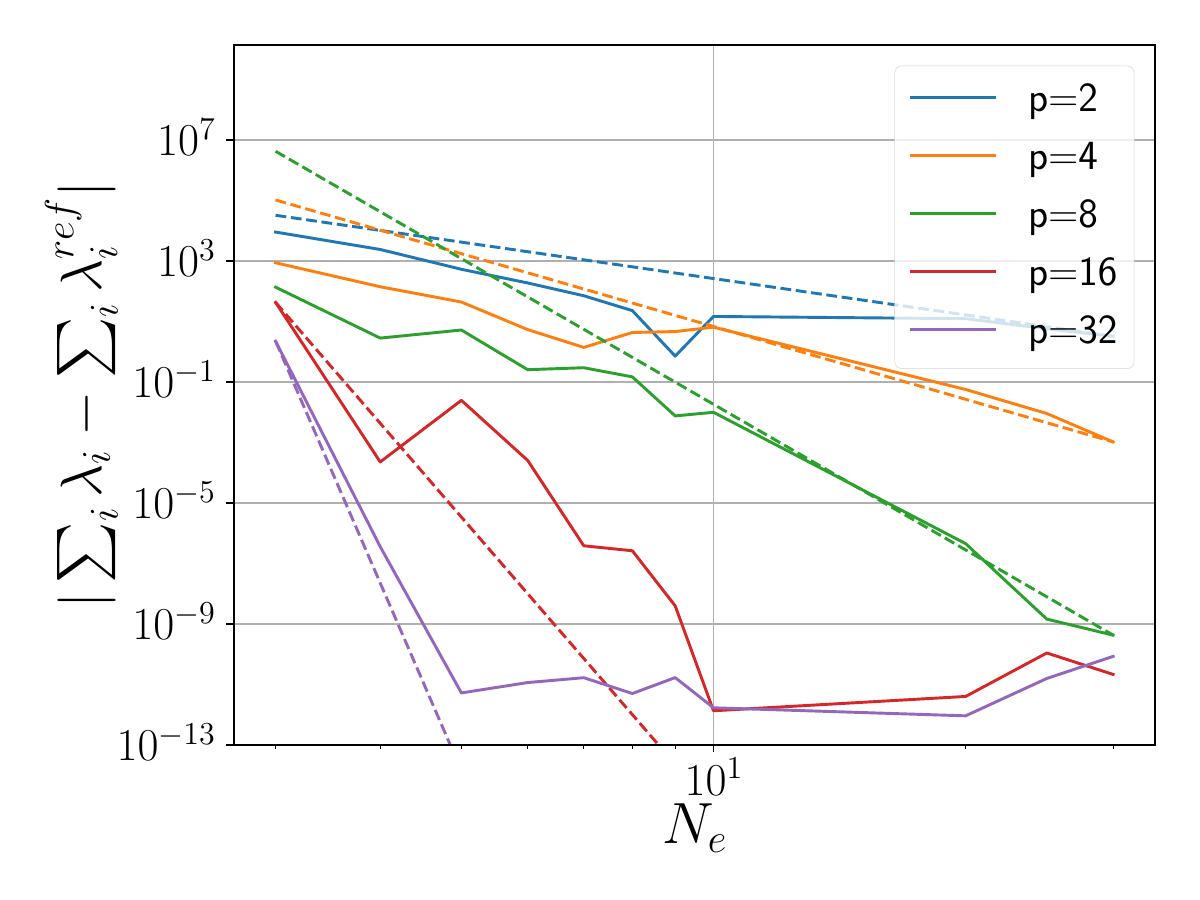}
}
\label{fig:harmonic_schroed_all}
\caption{Convergence studies for the Schrödinger equation with a harmonic oscillator potential.}
\end{figure}

Figure~\ref{fig:harmonic_schroed_conv} shows the total Schrödinger energy error with respect to the
polynomial order $p$ for various $N_{e}$ values. For fixed $N_{e}$, the
straight line on the log-linear graph shows the exponential dependency of the
error on $p$ until hitting the numerical precision limit (~$10^{-10}$).

The effect of $r_\mathrm{max}$ on the error was tested with five
elements. Figure~\ref{fig:harmonic_schroed_rmax} indicates that
$r_\mathrm{max} \ge 10$ gives a total energy error of $10^{-10}$ or lower.
Figure~\ref{fig:harmonic_schroed_rmax_eig} shows that the eigenvalues converge
to $10^{-11}$ under the same condition.

The dependence of the error on $N_{e}$, together with the
theoretical convergence rate ($N_{e}^{-2p}$), is shown in
Figure~\ref{fig:harmonic_schroed_ne}. The slope of the solid lines confirms that
the theoretical convergence rate is achieved. It is asymptotically achieved for
$p\le8$, while for $p>8$, it approaches the limiting numerical precision of
approximately $10^{-9}$.

For the Dirac equation, comparisons are with respect to \texttt{dftatom} results.

\begin{figure}
\centering
\subfloat[$p$ study for sum of eigenvalues]{
\label{fig:harmonic_dirac_conv}
\includegraphics[width=0.5\linewidth]{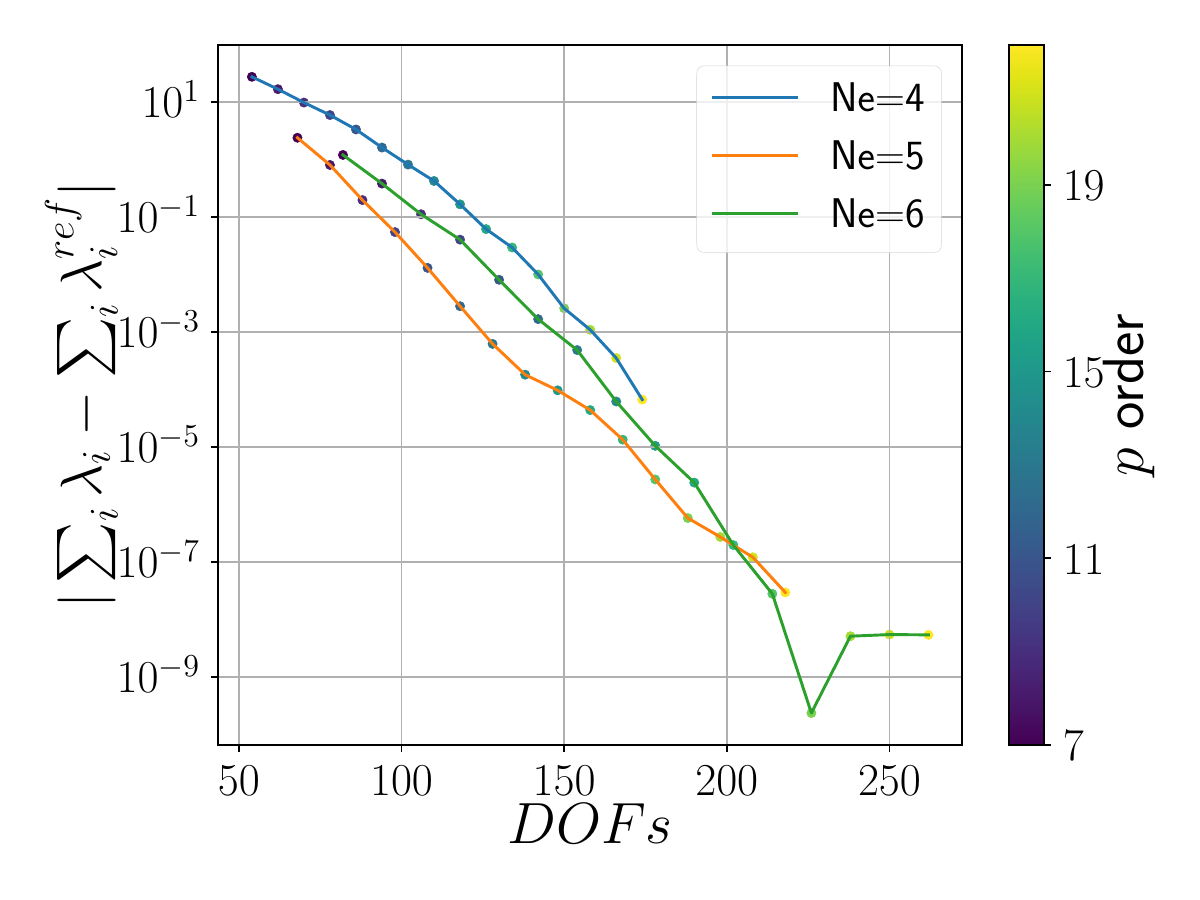}
}
\subfloat[$r_\mathrm{max}$ study for sum of eigenvalues]{
\label{fig:harmonic_dirac_rmax}
\includegraphics[width=0.5\linewidth]{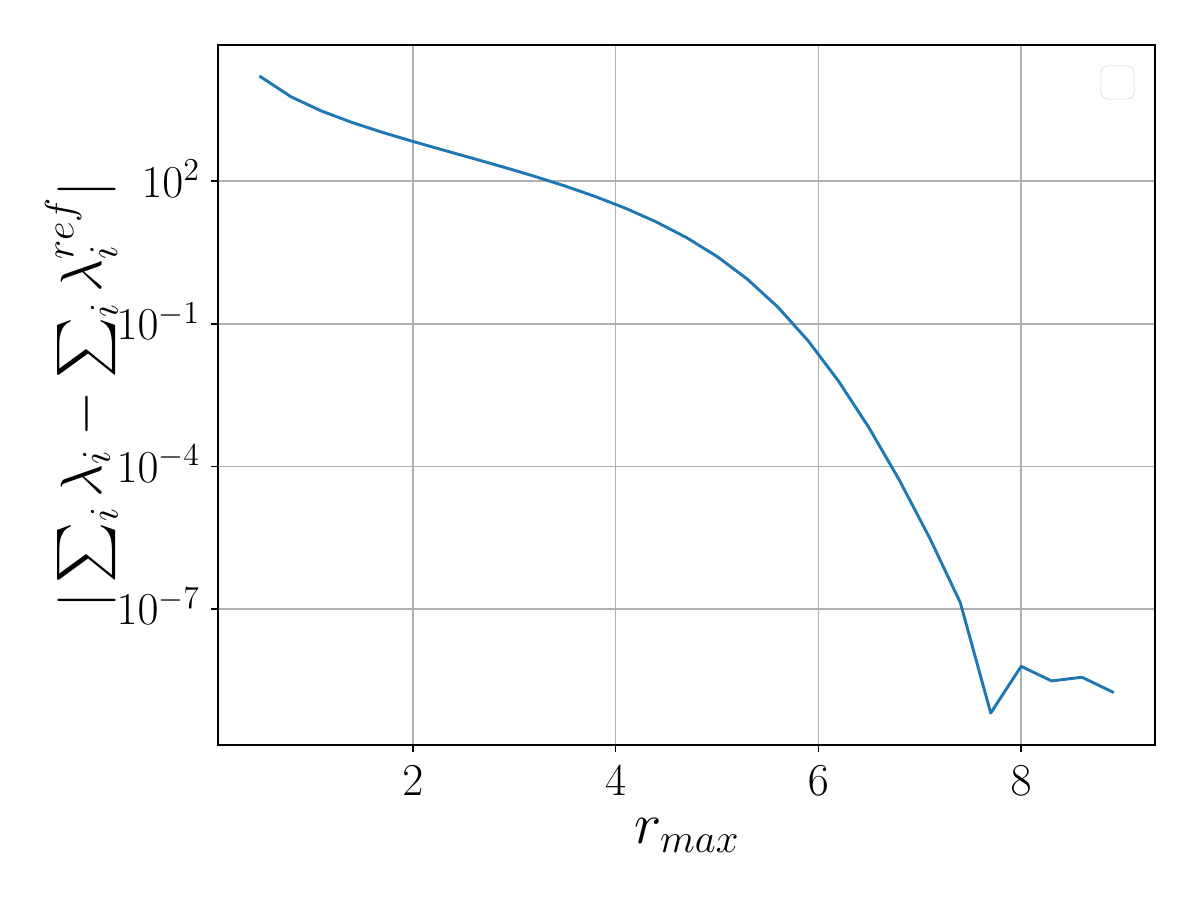}
}

\subfloat[$r_\mathrm{max}$ study for eigenvalues]{
\label{fig:harmonic_dirac_rmax_eig}
\includegraphics[width=0.5\linewidth]{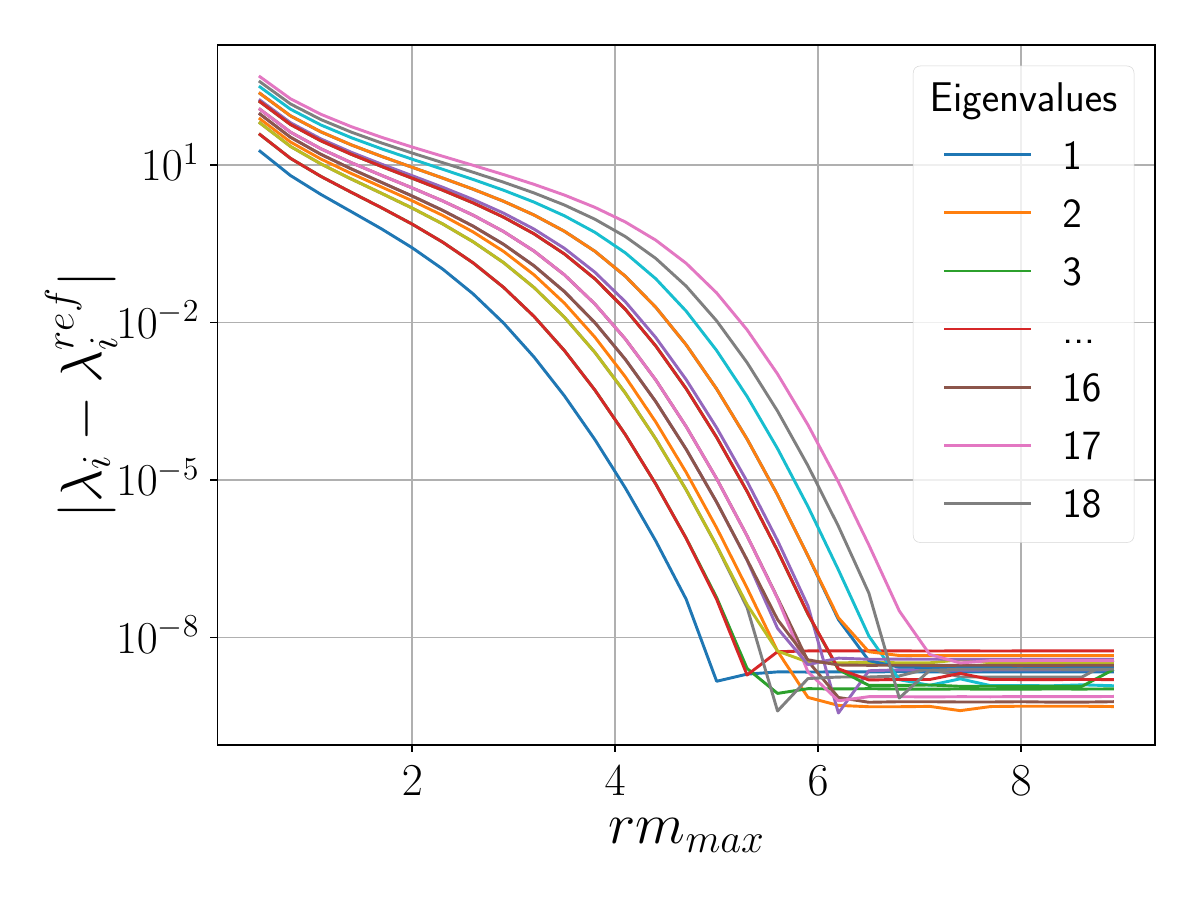}
}
\subfloat[$N_{e}$ study for sum of eigenvalues]{
\label{fig:harmonic_dirac_ne}
\includegraphics[width=0.5\linewidth]{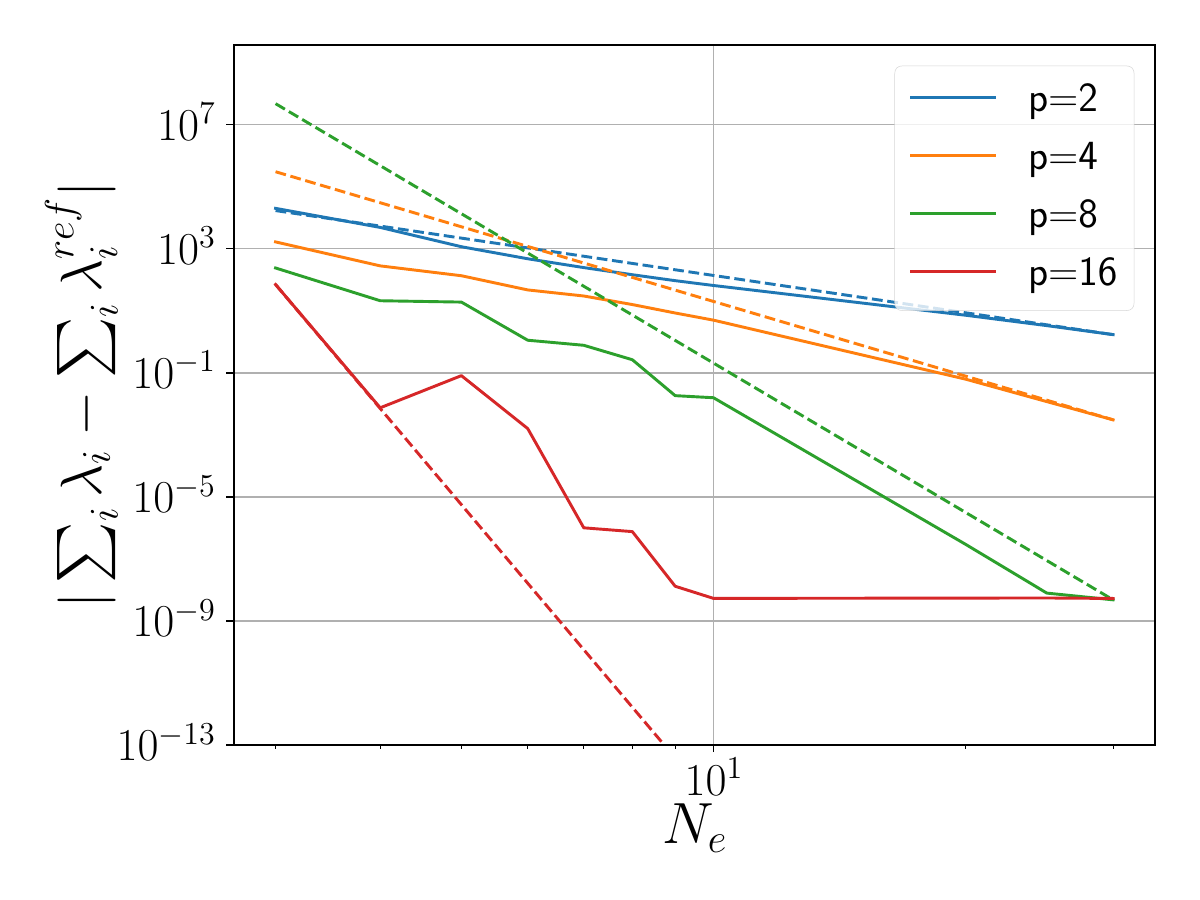}
}
\label{fig:harmonic_dirac_all}
\caption{Convergence studies for the Dirac equation with a harmonic oscillator potential.}
\end{figure}

Figure~\ref{fig:harmonic_dirac_conv} shows an exponential error decrease with respect to $p$
for various $N_{e}$. A similar pattern is observed with $r_\mathrm{max} \ge 10$,
reducing the total Dirac energy error to $10^{-8}$ or less
(Figure~\ref{fig:harmonic_dirac_rmax}). Eigenvalue convergence to $10^{-8}$ is
also seen (Figure~\ref{fig:harmonic_dirac_rmax_eig}).

The theoretical convergence rate ($N_{e}^{-2p}$) is confirmed in
Figure~\ref{fig:harmonic_dirac_ne}, attained asymptotically for $p\le8$ and reaching a
numerical precision limit of ~$10^{-9}$ for $p>8$.

\subsection{DFT calculations}
The accuracy of the DFT solvers using both Schrödinger and Dirac equations is
compared against \texttt{dftatom} results for the challenging case of uranium. The
non-relativistic Schrödinger calculation yields the following electronic
configuration:
$$1s^2 2s^2 2p^6 3s^2 3p^6 3d^{10} 4s^2 4p^6 4d^{10} 4f^{14} 5s^2 5p^6 5d^{10} 5f^3 6s^2 6p^6 6d^1 7s^2.$$
With the Dirac solver, the $l$-shell occupation splits according to the
degeneracy of the $j=l+\frac{1}{2}$ and $j=l-\frac{1}{2}$ subshells.

\begin{figure}
\centering
\subfloat[$p$ study for total energy]{
\label{fig:dft_schroed_conv}
\includegraphics[width=0.5\linewidth]{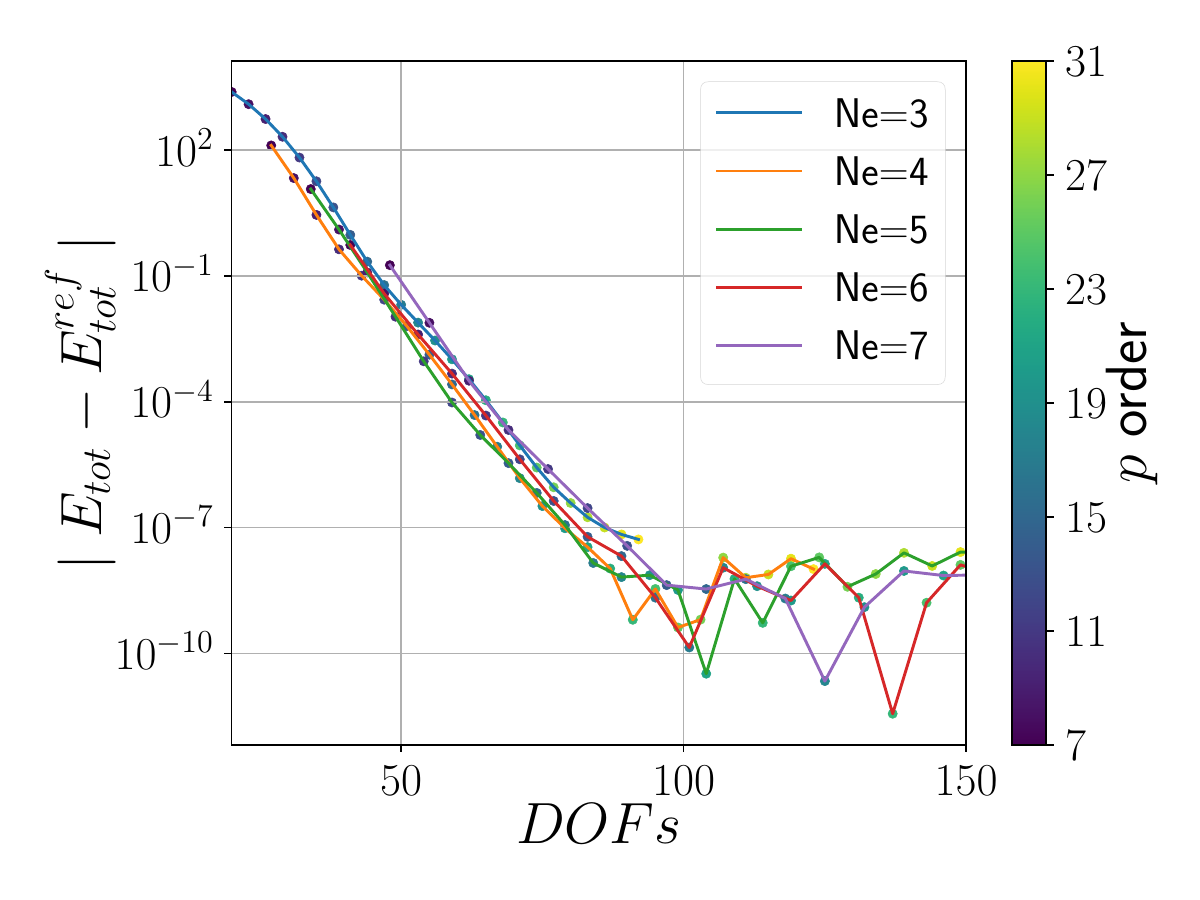}
}
\subfloat[$r_\mathrm{max}$ study for total energy]{
\label{fig:dft_schroed_rmax}
\includegraphics[width=0.5\linewidth]{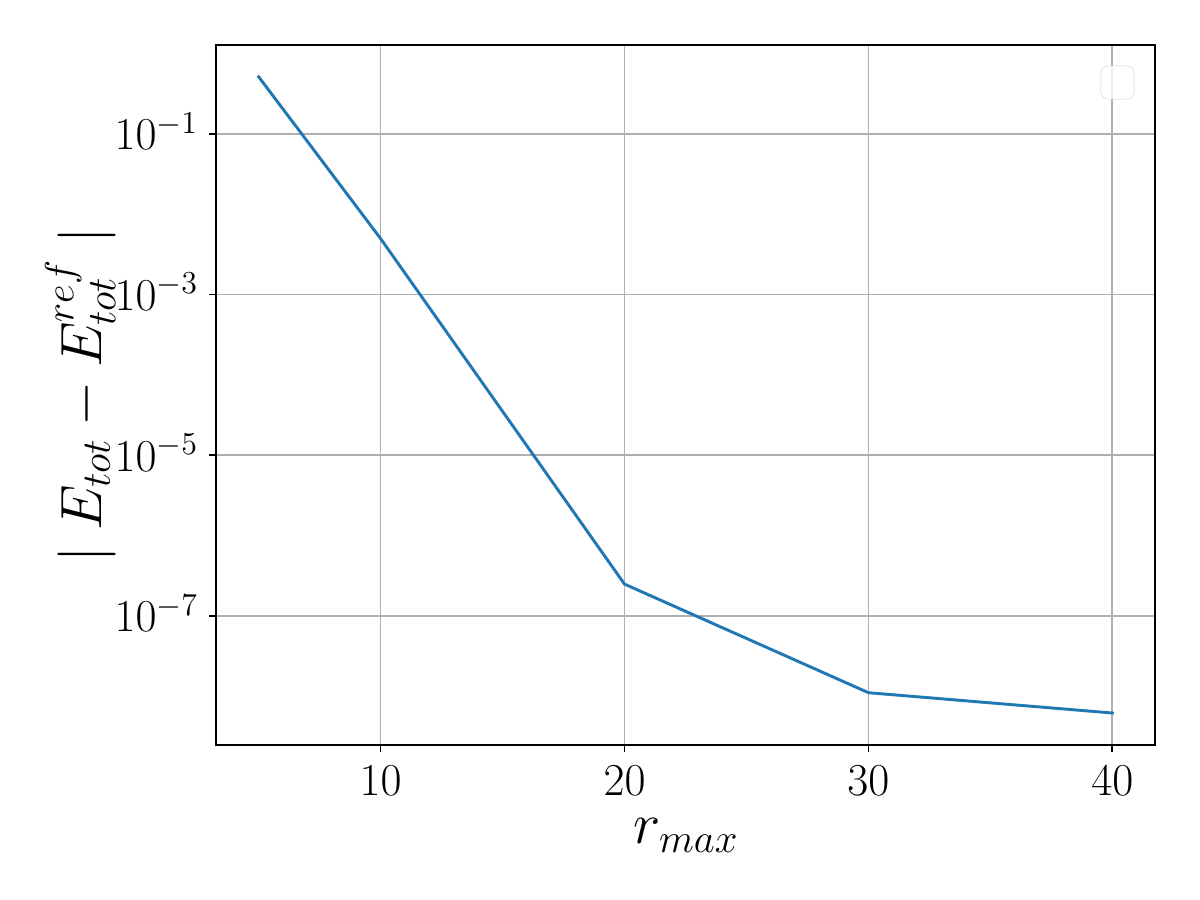}
}

\subfloat[$r_\mathrm{max}$ study for eigenvalues]{
\label{fig:dft_schroed_rmax_eig}
\includegraphics[width=0.5\linewidth]{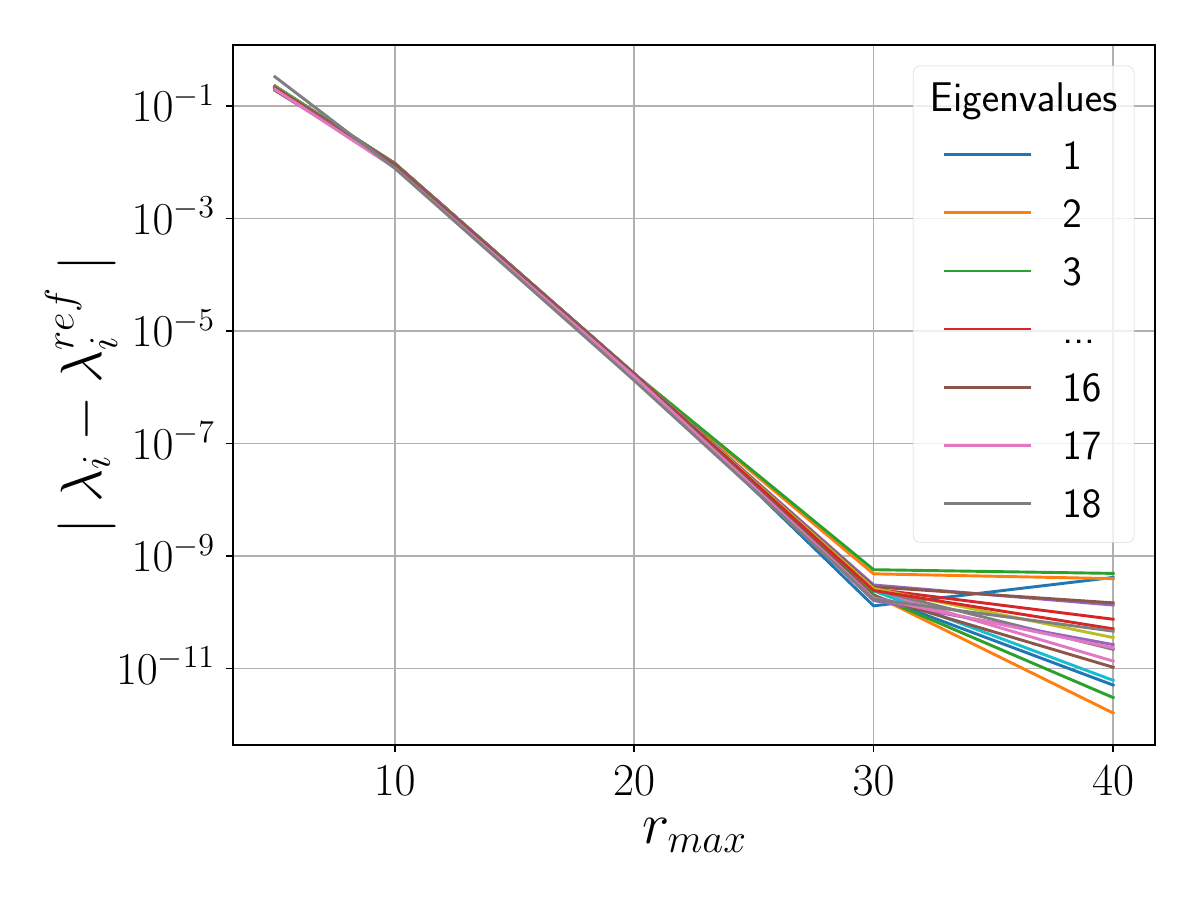}
}
\subfloat[$N_{e}$ study for total energy]{
\label{fig:dft_schroed_ne}
\includegraphics[width=0.5\linewidth]{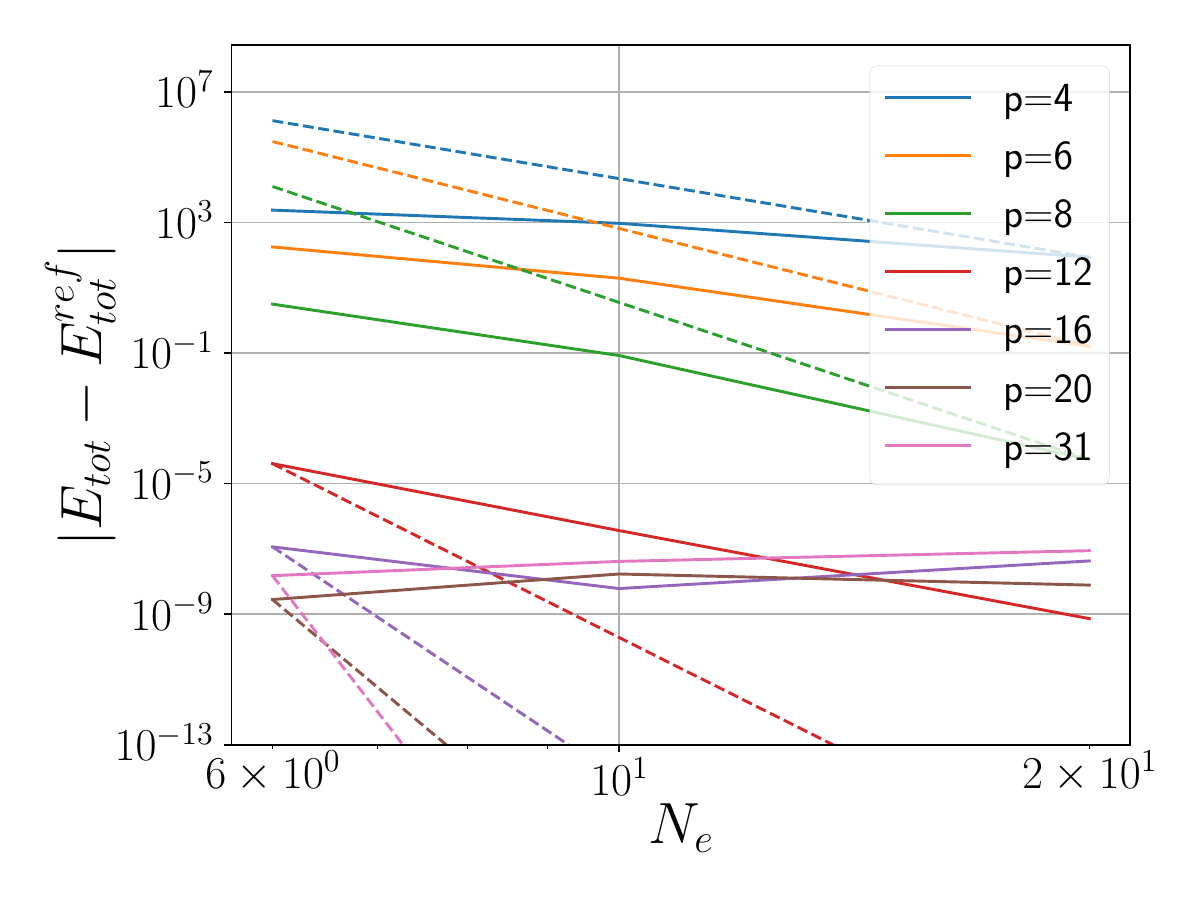}
}
\caption{Convergence studies for DFT with the Schrödinger equation for uranium.}
\end{figure}

\begin{figure}
\centering
\subfloat[$p$ study for total energy]{
\label{fig:dft_dirac_conv}
\includegraphics[width=0.5\linewidth]{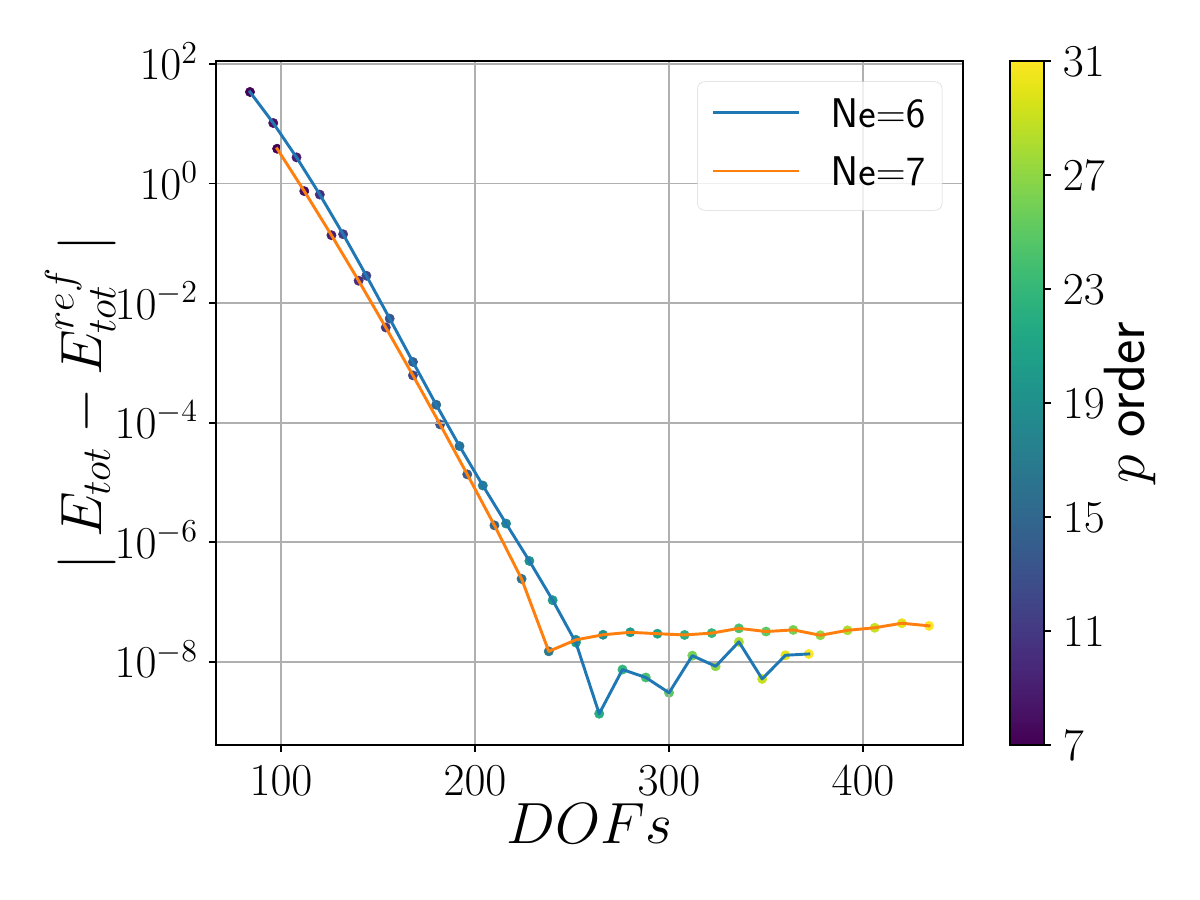}
}
\subfloat[$r_\mathrm{max}$ study for total energy]{
\label{fig:dft_dirac_rmax}
\includegraphics[width=0.5\linewidth]{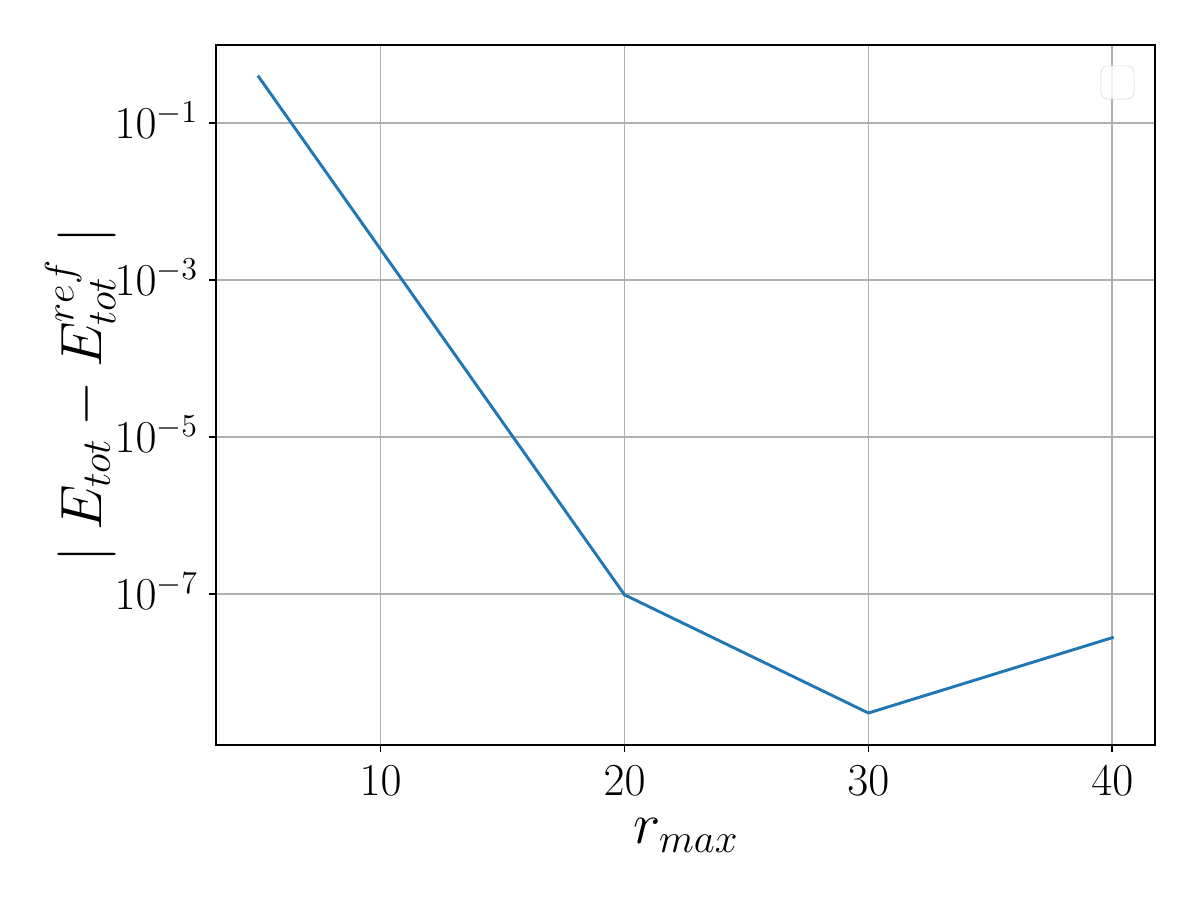}
}

\subfloat[$r_\mathrm{max}$ study for eigenvalues]{
\label{fig:dft_dirac_rmax_eig}
\includegraphics[width=0.5\linewidth]{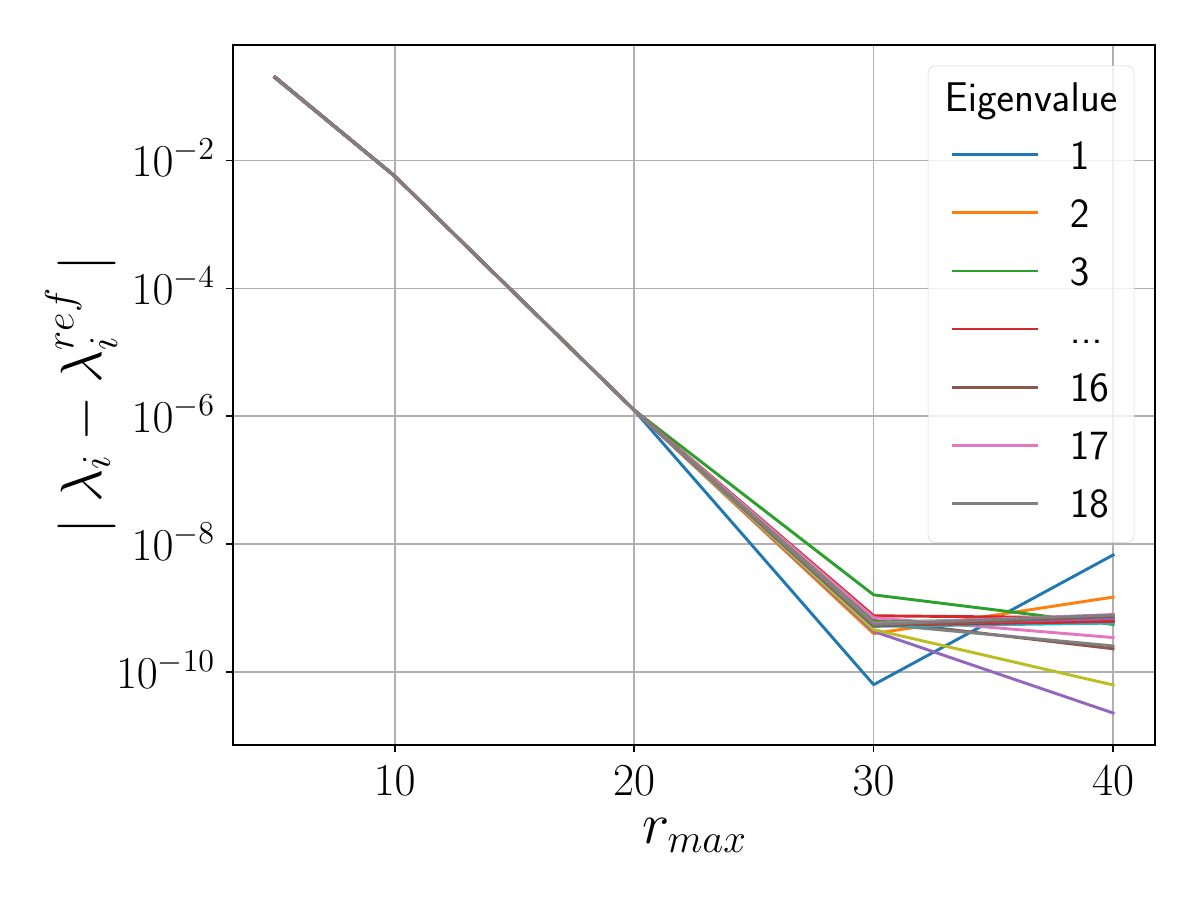}
}
\subfloat[$N_{e}$ study for total energy]{
\label{fig:dft_dirac_ne}
\includegraphics[width=0.5\linewidth]{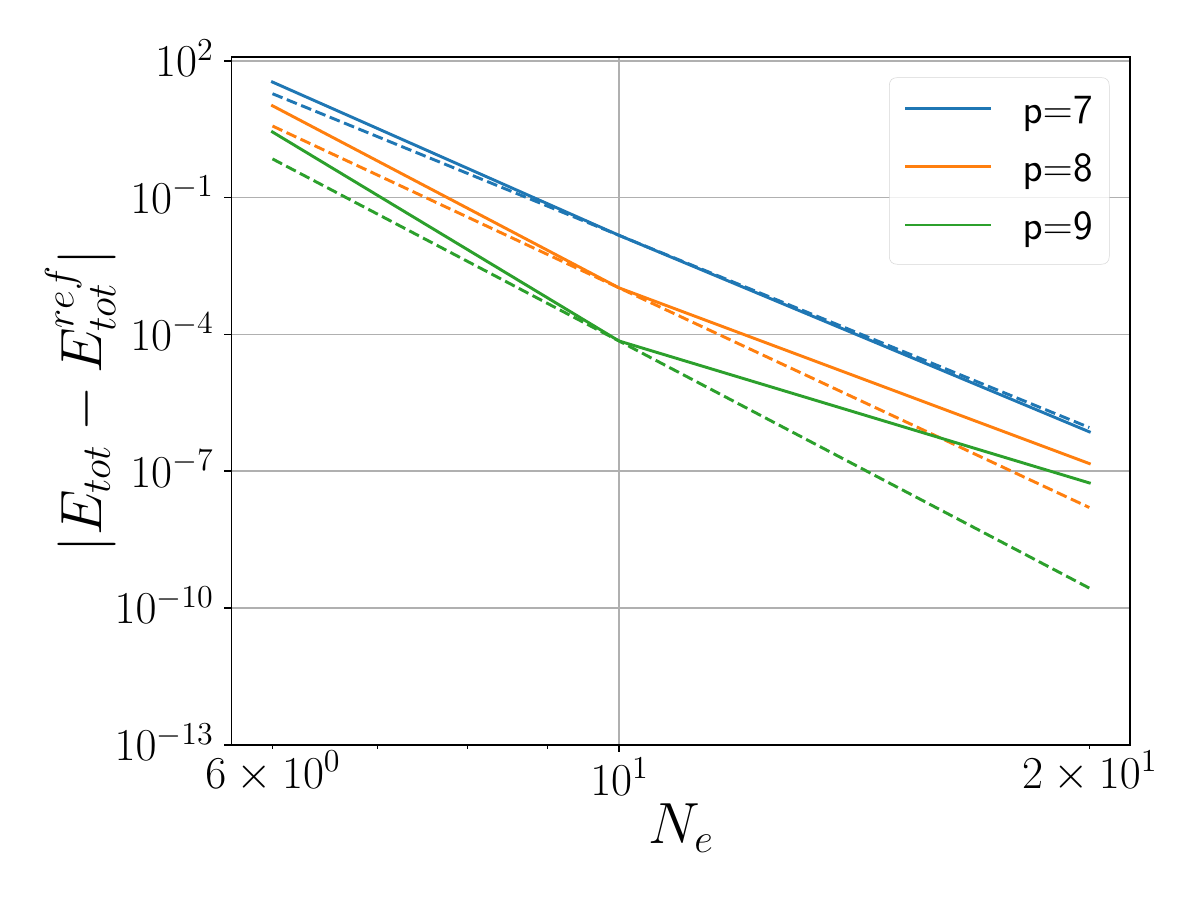}
}
\caption{Convergence studies for DFT with the Dirac equation for uranium.}
\end{figure}

Figures~\ref{fig:dft_schroed_conv} and \ref{fig:dft_dirac_conv} demonstrate an
exponential decrease in total energy error with increasing $p$ for various
$N_{e}$. As $r_\mathrm{max}$ increases to $\ge 25$ for five elements, the error
decreases to $10^{-8}$ or less (Figures~\ref{fig:dft_schroed_rmax} and
\ref{fig:dft_dirac_rmax}). Convergence of eigenvalues to $10^{-9}$ is seen at
$r_\mathrm{max} \ge 30$ (Figures~\ref{fig:dft_schroed_rmax_eig} and
\ref{fig:dft_dirac_rmax_eig}).

In Figures~\ref{fig:dft_schroed_ne} and \ref{fig:dft_dirac_ne}, the theoretical
convergence rate ($N_{e}^{-2p}$) manifests when $N_{e} \ge 12$, below which
numerical instabilities prevent convergence.

\subsection{Numerical considerations}\label{numericaleff}

The mesh parameters used to achieve $10^{-8}$ a.u. accuracy for the DFT
Schrödinger and Dirac calulations are shown in Table \ref{tab:mesh-params1}. The
mesh parameters used for $10^{-6}$ a.u. accuracy are shown in Table
\ref{tab:mesh-params2}.

\begin{table}[htbp]
  \centering
  \caption{Mesh parameters for achieving $10^{-8}$ a.u. accuracy in DFT Schrödinger and Dirac calculations of uranium.}
  \label{tab:mesh-params1}
  \begin{tabular}{ccc}
    \hline
    \textbf{Parameter} & \textbf{DFT Schrödinger} & \textbf{Dirac} \\
    \hline
    $Z$ & 92 & 92 \\
    $r_{\text{min}}$ & 0 & 0 \\
    $r_{\text{max}}$ & 50 & 30 \\
    $a$ & 200 & 600 \\
    $N_{e}$ & 4 & 6 \\
    $N_{q}$ & 53 & 64 \\
    $p$ & 26 & 25 \\
    $r_{0}$ & --- & 0.005 \\
    \hline
  \end{tabular}
\end{table}

\begin{table}[htbp]
  \centering
  \caption{Mesh parameters for achieving $10^{-6}$ a.u. accuracy in DFT Schrödinger and Dirac calculations of uranium.}
  \label{tab:mesh-params2}
  \begin{tabular}{ccc}
    \hline
    \textbf{Parameter} & \textbf{DFT Schrödinger} & \textbf{Dirac} \\
    \hline
    $Z$ & 92 & 92 \\
    $r_{\text{min}}$ & 0 & 0 \\
    $r_{\text{max}}$ & 30 & 30 \\
    $a$ & 200 & 100 \\
    $N_{e}$ & 4 & 5 \\
    $N_{q}$ & 35 & 40 \\
    $p$ & 17 & 24 \\
    $r_{0}$ & --- & 0.005 \\
    \hline
  \end{tabular}
\end{table}

\subsection{Benchmarks}

The presented \texttt{featom} implementation is written in Fortran and runs on
every platform with a modern Fortran compiler. To get an idea of the speed, we
benchmarked against \texttt{dftatom} on a laptop with an Apple M1 Max processor
using GFortran 11.3.0. We carry out the uranium DFT calculation to $10^{-6}$~a.u.
accuracy in total energy and all eigenvalues. The timings are as follows:

\begin{verbatim}
(Apple M1)  featom   dftatom
Schrödinger 28 ms     166 ms
Dirac       360 ms    276 ms
\end{verbatim}

We also benchmarked the Coulombic system from section 4.1:

\begin{verbatim}
(Apple M1)  featom   dftatom
Dirac        49 ms     64 ms
\end{verbatim}

\section{Summary and conclusions}\label{conclusions}

We have presented a robust and general finite element formulation for the
solution of the radial Schrödinger, Dirac, and Kohn--Sham equations of density
functional theory; and provided a modular, portable, and efficient Fortran
implementation,\texttt{featom}\footnote{Available on Github: \url{https://github.com/atomic-solvers/featom}}, along with interfaces to other languages and
full suite of examples and tests. To eliminate spurious states in the solution
of the Dirac equation, we work with the square of the Hamiltonian rather than
the Hamiltonian itself. Additionally, to eliminate convergence difficulties
associated with divergent derivatives and non-polynomial variation in the
vicinity of the origin, we solve for $\tilde P=\frac{P}{r^{\alpha}}$ and
$\tilde Q = \frac{Q}{r^{\alpha}}$ rather than for $P$ and $Q$ directly. We then employ a high-order finite element method to solve the resulting Schrödinger,
Dirac, and Poisson equations which can accommodate any potential, whether
singular Coulombic or finite, and any mesh, whether linear, exponential, or
otherwise. We have demonstrated the flexibility and accuracy of the associated
code with solutions of Schrödinger and Dirac equations for Coulombic and
harmonic oscillator potentials; and solutions of Kohn--Sham and Dirac--Kohn--Sham
equations for the challenging case of uranium, obtaining energies accurate to
$10^{-8}$ a.u., thus verifying current benchmarks
\cite{certikDftatomRobustGeneral2013,clarkAtomicReferenceData1997}. We have
shown detailed convergence studies in each case, providing mesh parameters to
facilitate straightforward convergence to any desired accuracy by simply
increasing the polynomial order.

At all points in the design of the associated code, we have tried to emphasize
simplicity and modularity so that the routines provided can be straightforwardly
employed for a range of applications purposes, while retaining high efficiency.
We have made the code available as open source to facilitate distribution,
modification, and use as needed. We expect the present solvers will be of
benefit to a broad range of large-scale electronic structure methods that rely on
atomic structure calculations and/or radial integration more generally as key
components.

\section{Acknowledgements}

We would like to thank Radek Kolman, Andreas Klöckner and Jed Brown for helpful
discussions. This work performed, in part, under the auspices of the U.S.
Department of Energy by Los Alamos National Laboratory under Contract
DE-AC52-06NA2539 and U.S. Department of Energy by Lawrence Livermore National
Laboratory under Contract DE-AC52-07NA27344. RG was partially supported by the
Icelandic Research Fund, grant number $217436052$.

\bibliography{../paper}

\appendix

\section{Derivations for the squared radial Dirac finite element formulation}\label{App:SqHamDerivDirac}
\subsection{Components of the Squared Radial Dirac Hamiltonian}\label{App:SqHamDeriv}

Starting from~(\ref{eq:SqasymH}), we have $H'$,

\begin{align*}
    H' = r^{2\alpha}\begin{pmatrix}
        V(r) + c^2 & c \left(-{\dv{r}}+{\left(\kappa-\alpha\right)\over r}\right) \\
        c \left({\dv{r}}+{\left(\kappa+\alpha\right)\over r}\right) & V(r) - c^2 \\
        \end{pmatrix}^2.
\end{align*}

Let
\begin{equation}
r^{-2\alpha} H'= \begin{pmatrix}
    G^{11} & G^{12} \\
    G^{21} & G^{22} \\
    \end{pmatrix},
\end{equation}

\noindent where

\begin{subequations}
\begin{align}
    G^{11} &= \left(V(r) + c^2\right)^2 + c^2 \left(-{\dv{r}}+{\left(\kappa-\alpha\right)\over r}\right) \left({\dv{r}}+{\left(\kappa+\alpha\right)\over r}\right), \\
    G^{12} &= \left(V(r) + c^2\right) c \left(-{\dv{r}}+{\left(\kappa-\alpha\right)\over r}\right) + c \left(-{\dv{r}}+{\left(\kappa-\alpha\right)\over r}\right) \left(V(r) - c^2\right), \\
    G^{21} &= c \left({\dv{r}}+{\left(\kappa+\alpha\right)\over r}\right) \left(V(r) + c^2\right) + \left(V(r) - c^2\right) c \left({\dv{r}}+{\left(\kappa+\alpha\right)\over r}\right), \\
    G^{22} &= \left(V(r) - c^2\right)^2 + c^2 \left({\dv{r}}+{\left(\kappa+\alpha\right)\over r}\right) \left({-\dv{r}}+{\left(\kappa-\alpha\right)\over r}\right)
\end{align}
\end{subequations}

\noindent We now simplify each term to obtain

\begin{subequations}
\begin{align}\label{eq:g11f}
    G^{11}f &= \left(V(r) + c^2\right)^2 f + c^2 \left(-{\dv{r}}+{\left(\kappa-\alpha\right)\over r}\right) \left({\dv{r}}+{\left(\kappa+\alpha\right)\over r}\right) f, \\
    G^{11}f &= \left(V(r) + c^2\right)^2 f + c^2 \left(-{\dv[2]{f}{r}}+{\left(\kappa-\alpha\right)\over r} {\dv{f}{r}}+ {\left(\kappa^2-\alpha^2\right) f\over r^2} -{\dv{r}}{\left(\kappa+\alpha\right)f\over r}\right).
\end{align}
\end{subequations}

\noindent Recall that

\begin{subequations}
\begin{align}\label{eq:derviIdent}
-{\dv{r}}{\left(\kappa+\alpha\right)f\over r} &= -\left(\kappa+\alpha\right) \left( {1 \over r}{\dv{f}{r}} - {f \over r^2} \right), \\
-{\dv{r}}{\left(\kappa+\alpha\right)f\over r} &= -{\left(\kappa+\alpha\right) \over r}{\dv{f}{r}} + {\left(\kappa+\alpha\right)f \over r^2}.
\end{align}
\end{subequations}

\noindent Substituting~(\ref{eq:derviIdent}) in~(\ref{eq:g11f}), with $\displaystyle\Phi=\frac{(\kappa(\kappa + 1) - \alpha(\alpha - 1))}{r^{2}}$ we have

\begin{align}
   G^{11}f &= \left(V(r) + c^2\right)^2 f + c^2 \left(-{\dv[2]{f}{r}}+{\left(\kappa-\alpha\right)\over r} {\dv{f}{r}}+\,{\left(\kappa^2-\alpha^2\right) f\over r^2} \right. \nonumber\\
           &\qquad\qquad\qquad\qquad\qquad\quad\left.-{\left(\kappa+\alpha\right) \over r}{\dv{f}{r}} + {\left(\kappa+\alpha\right)f \over r^2} \right)\\
           &= \left(V(r) + c^2\right)^2 f + c^2 \left(-{\dv[2]{f}{r}}-{2\alpha\over r} {\dv{f}{r}}+ \Phi f\right),
\end{align}

\noindent So we obtain
\begin{equation}
    G^{11}  = \left(V(r) + c^2\right)^2 + c^2 \left(-{\dv[2]{r}}-{2\alpha\over r} {\dv{r}}+ \Phi\right).
\end{equation}

\noindent For $G^{22}$, we have

\begin{align}\label{eq:g22}
    G^{22} &= \left(V(r) - c^2\right)^2 + c^2 \left({\dv{r}}+{\left(\kappa+\alpha\right)\over r}\right) \left({-\dv{r}}+{\left(\kappa-\alpha\right)\over r}\right) \nonumber\\
    &= \left(V(r) - c^2\right)^2 + c^2 \left({-\dv{r}}+{\left(-\kappa-\alpha\right)\over r}\right) \left({\dv{r}}+{\left(-\kappa+\alpha\right)\over r}\right).
\end{align}

\noindent Since

\begin{subequations}
\begin{align}\label{eq:g22deriv}
  \left(-{\dv{r}}+{\left(\kappa-\alpha\right)\over r}\right) \left({\dv{r}}+{\left(\kappa+\alpha\right)\over r}\right) &= \left(-{\dv[2]{r}}-{2\alpha\over r} {\dv{r}}\right.\nonumber\\
  &\qquad\left.+ \frac{(\kappa(\kappa + 1) - \alpha(\alpha - 1))}{r^{2}}\right),\\
  \left(-{\dv{r}}+{\left(-\kappa-\alpha\right)\over r}\right) \left({\dv{r}}+{\left(-\kappa+\alpha\right)\over r}\right) &= \left(-{\dv[2]{r}}-{2\alpha\over r} {\dv{r}}\right.\nonumber\\
  &\qquad\left.+ {\left(-\kappa\left(-\kappa+1\right)-\alpha\left(\alpha-1\right)\right) \over r^2}\right), \\
                                                                                                   &= \left(-{\dv[2]{r}}-{2\alpha\over r} {\dv{r}}\right.\nonumber\\
  &\qquad\left.+ {\left(\kappa\left(\kappa-1\right)-\alpha\left(\alpha-1\right)\right) \over r^2}\right).
\end{align}
\end{subequations}

\noindent Therefore, after substituting~(\ref{eq:g22deriv}) in~(\ref{eq:g22})

\begin{equation}
    G^{22} = \left(V(r) - c^2\right)^2 + c^2 \left(-{\dv[2]{r}}-{2\alpha\over r} {\dv{r}}+ {\left(\kappa\left(\kappa-1\right)-\alpha\left(\alpha-1\right)\right) \over r^2}\right).
\end{equation}

\noindent For $G^{12}$, we have

\begin{equation}
\begin{aligned}
    G^{12} f &= \left(V(r) + c^2\right) c \left(-{\dv{r}}+{\left(\kappa-\alpha\right)\over r}\right)f + c \left(-{\dv{r}}+{\left(\kappa-\alpha\right)\over r}\right) \left(V(r) - c^2\right)f, \\
                &= c{\left(\kappa-\alpha\right)\over r} \left(V(r) + c^2 + V(r) - c^2\right)f - \left(V(r) + c^2\right) c {\dv{f}{r}} -c {\dv{\left(V(r) - c^2\right)f}{r}}, \\
                &= c\left(2{\left(\kappa-\alpha\right)\over r} V(r)f - \left(V(r) + c^2\right) {\dv{f}{r}} - \left(V(r) - c^2\right){\dv{f}{r}} - V'(r)f \right), \\
                &= c\left(2{\left(\kappa-\alpha\right)\over r} V(r)f - 2V(r) {\dv{f}{r}} - V'(r)f \right),
\end{aligned}
\end{equation}

\noindent so

\begin{equation}
    G^{12} = c\left(2{\left(\kappa-\alpha\right)\over r} V(r) - 2V(r) {\dv{r}} - V'(r) \right).
\end{equation}

\noindent Similarly

\begin{equation}
\begin{aligned}
    G^{21}f &= c \left({\dv{r}}+{\left(\kappa+\alpha\right)\over r}\right) \left(V(r) + c^2\right)f + \left(V(r) - c^2\right) c \left({\dv{r}}+{\left(\kappa+\alpha\right)\over r}\right)f \\
                &= c{\left(\kappa+\alpha\right)\over r} \left(V(r) + c^2 + V(r) - c^2\right)f + \left(V(r) - c^2\right) c {\dv{f}{r}} +c {\dv{\left(V(r) + c^2\right)f}{r}} \\
                &= c\left(2{\left(\kappa+\alpha\right)\over r} V(r)f + \left(V(r) - c^2\right) {\dv{f}{r}} + \left(V(r) + c^2\right){\dv{f}{r}} + V'(r)f \right) \\
                &= c\left(2{\left(\kappa+\alpha\right)\over r} V(r)f + 2V(r) {\dv{f}{r}} + V'(r)f \right)
\end{aligned}
\end{equation}

\noindent which leads to

\begin{equation}
    G^{21} = c\left(2{\left(\kappa+\alpha\right)\over r} V(r) + 2V(r) {\dv{r}} + V'(r) \right).
\end{equation}

\subsection{Weak formulation of the squared Hamiltonian for the radial Dirac}\label{App:weakSqDH}

Starting from~(\ref{squared_fem1}),

\begin{equation*}
A =
    \int_0^\infty\!\!
\begin{pmatrix} \phi_i^a(r) & \phi_i^b(r) \\ \end{pmatrix}
    H'
        \begin{pmatrix}
        \phi_j^a(r) \\
        \phi_j^b(r) \\
        \end{pmatrix}
        \dd r, \qquad
A = \begin{pmatrix}
    A^{11} & A^{12} \\
    A^{21} & A^{22} \\
    \end{pmatrix}
\end{equation*}

\noindent and also

\begin{align*}
\phi_i^a(r)  &= \begin{cases}
                \pi_i(r) & \text{for $i=1, \dots, N$}, \\
                0        & \text{for $i=N+1, \dots, 2N$}.
            \end{cases} \\
\phi_i^b(r)  &= \begin{cases}
                0            & \text{for $i=1, \dots, N$}, \\
                \pi_{i-N}(r) & \text{for $i=N+1, \dots, 2N$}.
            \end{cases}
\end{align*}

With $\displaystyle\Phi=\frac{(\kappa(\kappa + 1) - \alpha(\alpha - 1))}{r^{2}}$ we have

\begin{subequations}
\begin{align}
  A^{11}_{ij} &= \int_{0}^{\infty}\pi_{i}(r)\left(  r^{2\alpha}(V(r) + c^{2})^{2} + r^{2\alpha}c^{2} \left( - \dv[2]{r} - \frac{2\alpha}{r}\dv{r} + \Phi \right) \right) \pi_{j}(r)\textrm{d} r,\label{eq:a11_1} \\
  A^{12}_{ij} &= \int_{0}^{\infty}\pi_{i}(r)r^{2\alpha}c\left( 2\frac{(\kappa - \alpha)}{r}V(r) - 2V(r)\dv{r} - V'(r) \right)\pi_{j}(r)\textrm{d} r, \\
  A_{ij}^{21} &= \int_{0}^{\infty}\pi_{i}(r)r^{2\alpha}c\left( 2\frac{(\kappa + \alpha)}{r}V(r) + 2V(r)\dv{r} + V'(r) \right)\pi_{j}(r)\textrm{d} r, \\
  A_{ij}^{22} &= \int_{0}^{\infty}\pi_{i}(r)\left( r^{2\alpha}( V(r) - c^{2} )^{2} + r^{2\alpha}c^{2} \left( -\dv[2]{r} - \frac{2\alpha}{r}\dv{r} + \Phi \right) \right)\textrm{d} r\label{eq:a22_1}.
\end{align}
\end{subequations}

The second term in (\ref{eq:a11_1}) and (\ref{eq:a22_1}) can be simplified as

\begin{subequations}
\begin{align}
\hspace{2em}&\hspace{-2em} \int_0^\infty \pi_i(r) r^{2\alpha}c^2\left(-{\dv[2]{r}}-{2\alpha\over r} {\dv{r}}\right)\pi_j(r)\textrm{d} r \\
            &= \int_0^\infty {\dv{\pi_j(r)}{r}} c^2 \left(r^{2\alpha}{\dv{\pi_i(r)}{r}} + 2\alpha r^{2\alpha-1} \pi_i(r) \right)\textrm{d} r\nonumber\\
  &\quad - \pi_i(r)\left(r^{2\alpha}c^2{2\alpha\over r} {\dv{r}}\right)\pi_j(r) - \cancelto{0}{\pi_i(r) r^{2\alpha} c^2 {\dv{\pi_j(r)}{r}} \Biggr|^{\infty}_0} \\
    &= \int_0^\infty {\dv{\pi_j(r)}{r}} c^2 r^{2\alpha}{\dv{\pi_i(r)}{r}}\textrm{d} r.
\end{align}
\end{subequations}

\noindent Therefore,
\begin{subequations}
\begin{align}
  \label{eq:dirbcApp}
    A^{11}_{ij} = \int_0^\infty r^{2\alpha} \left( \pi_i(r) \left(\left(V(r) + c^2\right)^2 + c^2 \Phi \right)\pi_j(r) + \pi_j'(r) c^2 \pi_i'(r)\right)\textrm{d} r \\
    A^{22}_{ij} = \int_0^\infty r^{2\alpha} \left( \pi_i(r) \left(\left(V(r) - c^2\right)^2 + c^2 \Phi \right)\pi_j(r) + \pi_j'(r) c^2 \pi_i'(r)\right)\textrm{d} r
\end{align}
\end{subequations}

\noindent which implies that $A^{11}$ and $A^{22}$ are symmetric.

\noindent The off diagonal terms can be simplified by rewriting the derivative of the
potential $V'$ using integration by parts

\begin{subequations}
\begin{align}
A^{12}_{ij} &= \int_0^\infty \pi_i(r)
    r^{2\alpha}c\left(2{\left(\kappa-\alpha\right)\over r} V(r) - 2V(r)
    {\dv{r}} - V'(r) \right) \pi_j(r) \textrm{d} r \\
            &= \int_0^\infty \pi_i(r) r^{2\alpha}c\left(
                2{\left(\kappa-\alpha\right)\over r} V(r)
                - 2V(r) {\dv{r}}
                \right) \pi_j(r)\textrm{d} r\nonumber \\
         &\quad + \int_0^\infty \pi_i(r) r^{2\alpha}c\left(
                - V'(r) \right) \pi_j(r)\textrm{d} r \\
            &= \int_0^\infty \pi_i(r) r^{2\alpha}c\left(
                2{\left(\kappa-\alpha\right)\over r} V(r)
                - 2V(r) {\dv{r}}
                \right) \pi_j(r)\textrm{d} r\nonumber \\
         &\quad + \int_0^\infty V(r) \left( \pi_i(r) r^{2\alpha}c \pi_j(r) \right)' {\textrm{d} r}
    + \cancelto{0}{V(r)\pi_i(r) r^{2\alpha} c \pi_j(r) \Biggr|^{\infty}_0}\\
            &= \int_0^\infty \pi_i(r) r^{2\alpha}c\left(
                2{\left(\kappa-\alpha\right)\over r} V(r)
                - 2V(r) {\dv{r}}
                \right) \pi_j(r)\textrm{d} r\nonumber \\
         &\quad + \int_0^\infty c V(r) \left(
                \pi_i'(r)\pi_j(r) + \pi_i(r)\pi_j'(r) + 2\pi_i\pi_j{\alpha\over
                r}
                \right) r^{2\alpha}\textrm{d} r  \\
            &= \int_0^\infty r^{2\alpha} c V(r) \left(
                    - \pi_i(r) \pi_j'(r)
                    + 2{\kappa\over r} \pi_i(r) \pi_j(r)
                    + \pi_i'(r) \pi_j(r)
                        \right)\textrm{d} r.
\end{align}
\end{subequations}
Similarly,
\begin{align}
A^{21}_{ij} &= \int_0^\infty \pi_i(r)
    r^{2\alpha}c\left(2{\left(\kappa+\alpha\right)\over r} V(r) + 2V(r)
    {\dv{r}} + V'(r) \right) \pi_j(r) \textrm{d} r \notag \\
            &= \int_0^\infty r^{2\alpha} c V(r) \left(
                    + \pi_i(r) \pi_j'(r)
                    + 2{\kappa\over r} \pi_i(r) \pi_j(r)
                    - \pi_i'(r) \pi_j(r)\right)\textrm{d} r
\end{align}

\noindent By exchanging the $ij$ indices, we see that $A^{12}_{ij} = A^{21}_{ji}$.
Since $A^{11}$ and $A^{22}$ are symmetric and $A^{12}_{ij} = A^{21}_{ji}$, the
finite element matrix $A$ is symmetric. The last four equations for $A^{11}$,
$A^{22}$, $A^{12}$ and $A^{21}$ are the equations that are implemented in the
code.

\section{Converged runs for systems}\label{App:SysRes}

The Coulomb potential results and the harmonic Schrödinger results are compared to the analytic results. The remaining systems are compared against \texttt{dftatom}~\cite{certikDftatomRobustGeneral2013}.

\subsection{Schrödinger equation with a Coulomb potential with $Z=92$}

\begin{verbatim}
  Z  rmax   Ne       a  p Nq DOFs
 92  50.0    7   100.0 31 53  216   -10972.971428571371

 Comparison of calculated and reference energies

 Total energy:
                   E               E_ref     error
 -10972.971428571371 -10972.971428571391  2.00E-11
 Eigenvalues:
   n                   E               E_ref     error
   1  -4231.999999999996  -4232.000000000000  3.64E-12
   2  -1057.999999999942  -1058.000000000000  5.80E-11
   3  -1057.999999999930  -1058.000000000000  7.03E-11
   4   -470.222222222330   -470.222222222222  1.08E-10
   5   -470.222222222260   -470.222222222222  3.74E-11
   6   -470.222222222187   -470.222222222222  3.56E-11
   7   -264.500000000015   -264.500000000000  1.49E-11
   8   -264.499999999977   -264.500000000000  2.28E-11
   9   -264.499999999988   -264.500000000000  1.24E-11
  10   -264.499999999965   -264.500000000000  3.52E-11
  11   -169.280000000003   -169.280000000000  3.27E-12
  12   -169.279999999990   -169.280000000000  9.66E-12
  13   -169.279999999997   -169.280000000000  2.96E-12
  14   -169.280000000012   -169.280000000000  1.20E-11
  15   -169.279999999994   -169.280000000000  5.68E-12
  16   -117.555555555556   -117.555555555556  7.53E-13
  17   -117.555555555550   -117.555555555556  5.37E-12
  18   -117.555555555554   -117.555555555556  1.71E-12
  19   -117.555555555564   -117.555555555556  8.38E-12
  20   -117.555555555555   -117.555555555556  4.26E-13
  21   -117.555555555565   -117.555555555556  9.02E-12
  22    -86.367346938776    -86.367346938770  6.45E-12
  23    -86.367346938773    -86.367346938770  3.27E-12
  24    -86.367346938773    -86.367346938770  3.40E-12
  25    -86.367346938780    -86.367346938770  1.02E-11
  26    -86.367346938780    -86.367346938770  9.54E-12
  27    -86.367346938787    -86.367346938770  1.73E-11
  28    -86.367346938776    -86.367346938770  6.38E-12
\end{verbatim}

\subsection{Dirac equation with a Coulomb potential with $Z=92$}

\begin{verbatim}
 Comparison of calculated and reference energies

 Total energy:
                   E               E_ref     error
 -16991.208873101066 -16991.208873101074  7.28E-12
 Eigenvalues:
   n                   E               E_ref     error
   1  -4861.198023119523  -4861.198023119372  1.51E-10
   2  -1257.395890257783  -1257.395890257889  1.06E-10
   3  -1089.611420919755  -1089.611420919875  1.20E-10
   4  -1257.395890257871  -1257.395890257889  1.82E-11
   5   -539.093341793807   -539.093341793890  8.37E-11
   6   -489.037087678178   -489.037087678200  2.18E-11
   7   -539.093341793909   -539.093341793890  1.82E-11
   8   -476.261595161184   -476.261595161155  2.91E-11
   9   -489.037087678164   -489.037087678200  3.64E-11
  10   -295.257844100346   -295.257844100397  5.09E-11
  11   -274.407758840011   -274.407758840065  5.46E-11
  12   -295.257844100397   -295.257844100397  0.00E+00
  13   -268.965877827151   -268.965877827130  2.18E-11
  14   -274.407758840043   -274.407758840065  2.18E-11
  15   -266.389447187838   -266.389447187816  2.18E-11
  16   -268.965877827159   -268.965877827130  2.91E-11
  17   -185.485191678526   -185.485191678552  2.55E-11
  18   -174.944613583455   -174.944613583462  7.28E-12
  19   -185.485191678534   -185.485191678552  1.82E-11
  20   -172.155252323719   -172.155252323737  1.82E-11
  21   -174.944613583462   -174.944613583462  0.00E+00
  22   -170.828937049882   -170.828937049879  3.64E-12
  23   -172.155252323751   -172.155252323737  1.46E-11
  24   -170.049934288658   -170.049934288552  1.06E-10
  25   -170.828937049879   -170.828937049879  0.00E+00
  26   -127.093638842631   -127.093638842631  0.00E+00
  27   -121.057538029541   -121.057538029549  7.28E-12
  28   -127.093638842609   -127.093638842631  2.18E-11
  29   -119.445271987144   -119.445271987141  3.64E-12
  30   -121.057538029545   -121.057538029549  3.64E-12
  31   -118.676410324362   -118.676410324351  1.09E-11
  32   -119.445271987144   -119.445271987141  3.64E-12
  33   -118.224144624910   -118.224144624903  7.28E-12
  34   -118.676410324355   -118.676410324351  3.64E-12
  35   -117.925825597409   -117.925825597293  1.16E-10
  36   -118.224144624906   -118.224144624903  3.64E-12
  37    -92.440787600957    -92.440787600943  1.46E-11
  38    -88.671749052020    -88.671749052017  3.64E-12
  39    -92.440787600932    -92.440787600943  1.09E-11
  40    -87.658287631897    -87.658287631893  3.64E-12
  41    -88.671749052013    -88.671749052017  3.64E-12
  42    -87.173966671959    -87.173966671948  1.09E-11
  43    -87.658287631897    -87.658287631893  3.64E-12
  44    -86.888766390952    -86.888766390941  1.09E-11
  45    -87.173966671948    -87.173966671948  0.00E+00
  46    -86.700519572882    -86.700519572809  7.28E-11
  47    -86.888766390948    -86.888766390941  7.28E-12
  48    -86.566875102300    -86.566875102359  5.82E-11
  49    -86.700519572820    -86.700519572809  1.09E-11
\end{verbatim}

\subsection{Schrödinger equation with a harmonic oscillator potential}
\begin{verbatim}
  Z  rmax   Ne       a  p Nq DOFs
 92  50.0    7   100.0 31 64  216      209.999999999991

 Comparison of calculated and reference energies

 Total energy:
                   E               E_ref     error
    209.999999999991    210.000000000000  9.35E-12
 Eigenvalues:
   n                   E               E_ref     error
   1      1.500000000000      1.500000000000  3.41E-13
   2      3.499999999999      3.500000000000  1.43E-12
   3      2.500000000000      2.500000000000  3.60E-14
   4      5.499999999999      5.500000000000  1.27E-12
   5      4.500000000000      4.500000000000  1.22E-13
   6      3.500000000000      3.500000000000  3.87E-13
   7      7.499999999998      7.500000000000  1.59E-12
   8      6.500000000000      6.500000000000  6.84E-14
   9      5.500000000000      5.500000000000  9.41E-14
  10      4.500000000000      4.500000000000  3.95E-13
  11      9.499999999998      9.500000000000  1.71E-12
  12      8.500000000000      8.500000000000  4.33E-13
  13      7.500000000000      7.500000000000  1.64E-13
  14      6.500000000000      6.500000000000  2.39E-13
  15      5.500000000000      5.500000000000  1.23E-13
  16     11.499999999998     11.500000000000  1.68E-12
  17     10.499999999999     10.500000000000  6.34E-13
  18      9.500000000000      9.500000000000  2.36E-13
  19      8.500000000000      8.500000000000  8.70E-14
  20      7.500000000000      7.500000000000  1.68E-13
  21      6.500000000000      6.500000000000  5.24E-14
  22     13.499999999998     13.500000000000  1.64E-12
  23     12.500000000000     12.500000000000  3.57E-13
  24     11.500000000000     11.500000000000  1.30E-13
  25     10.500000000000     10.500000000000  2.38E-13
  26      9.500000000000      9.500000000000  1.07E-14
  27      8.500000000000      8.500000000000  1.90E-13
  28      7.500000000000      7.500000000000  2.22E-14
\end{verbatim}

\subsection{Dirac equation with a harmonic oscillator potential}
\begin{verbatim}
  Z  rmax   Ne       a  p Nq DOFs
 92  50.0    7   100.0 23 53  320      367.470826694655

 Comparison of calculated and reference energies

 Total energy:
                   E               E_ref     error
    367.470826694655    367.470826700800  6.15E-09
 Eigenvalues:
   n              E          E_ref     error
   1     1.49999501     1.49999501  3.46E-11
   2     3.49989517     3.49989517  8.93E-12
   3     2.49997504     2.49997504  5.07E-11
   4     2.49993510     2.49993511  6.83E-10
   5     5.49971548     5.49971548  3.35E-11
   6     4.49983527     4.49983527  1.15E-12
   7     4.49979534     4.49979534  3.15E-09
   8     3.49994176     3.49994176  2.21E-11
   9     3.49987520     3.49987520  2.43E-11
  10     7.49945594     7.49945594  8.47E-11
  11     6.49961565     6.49961565  8.60E-12
  12     6.49957572     6.49957572  3.51E-10
  13     5.49976206     5.49976206  1.70E-11
  14     5.49969551     5.49969551  3.34E-12
  15     4.49989517     4.49989517  1.82E-11
  16     4.49980199     4.49980199  1.92E-11
  17     9.49911657     9.49911657  1.39E-10
  18     8.49931620     8.49931620  9.65E-11
  19     8.49927627     8.49927627  3.27E-10
  20     7.49950252     7.49950252  8.64E-11
  21     7.49943598     7.49943598  6.13E-11
  22     6.49967554     6.49967554  1.81E-11
  23     6.49958238     6.49958238  1.16E-10
  24     5.49983526     5.49983526  3.35E-11
  25     5.49971547     5.49971547  1.40E-11
  26    11.49869739    11.49869739  1.91E-10
  27    10.49893692    10.49893692  1.04E-10
  28    10.49889700    10.49889700  2.10E-10
  29     9.49916315     9.49916315  1.59E-10
  30     9.49909661     9.49909661  1.22E-10
  31     8.49937608     8.49937608  7.65E-11
  32     8.49928292     8.49928292  1.31E-10
  33     7.49957572     7.49957572  7.69E-11
  34     7.49945594     7.49945594  6.51E-11
  35     6.49976205     6.49976205  1.62E-12
  36     6.49961565     6.49961565  4.14E-11
  37    13.49819839    13.49819839  2.13E-10
  38    12.49847782    12.49847782  1.97E-10
  39    12.49843791    12.49843790  6.08E-10
  40    11.49874396    11.49874396  1.25E-10
  41    11.49867742    11.49867742  1.74E-10
  42    10.49899680    10.49899680  1.30E-10
  43    10.49890365    10.49890365  1.45E-10
  44     9.49923634     9.49923634  1.23E-10
  45     9.49911657     9.49911657  1.47E-10
  46     8.49946258     8.49946258  5.86E-11
  47     8.49931619     8.49931619  1.05E-10
  48     7.49967553     7.49967553  5.61E-11
  49     7.49950251     7.49950251  5.40E-11
\end{verbatim}

\subsection{DFT with the Schrödinger equation for uranium}
\begin{verbatim}
 Total energy:
               E           E_ref     error
 -25658.41788885 -25658.41788885  3.17E-09

 Eigenvalues:
   n               E           E_ref     error
   1  -3689.35513984  -3689.35513984  7.72E-10
   2   -639.77872809   -639.77872809  3.69E-10
   3   -619.10855018   -619.10855018  3.77E-10
   4   -161.11807321   -161.11807321  1.40E-11
   5   -150.97898016   -150.97898016  3.23E-11
   6   -131.97735828   -131.97735828  1.87E-10
   7    -40.52808425    -40.52808425  4.94E-11
   8    -35.85332083    -35.85332083  6.29E-11
   9    -27.12321230    -27.12321230  8.18E-11
  10    -15.02746007    -15.02746007  1.07E-10
  11     -8.82408940     -8.82408940  5.36E-11
  12     -7.01809220     -7.01809220  5.26E-11
  13     -3.86617513     -3.86617513  5.65E-11
  14     -0.36654335     -0.36654335  1.05E-11
  15     -1.32597632     -1.32597632  4.04E-11
  16     -0.82253797     -0.82253797  2.32E-12
  17     -0.14319018     -0.14319018  1.42E-11
  18     -0.13094786     -0.13094786  5.33E-11
\end{verbatim}

\subsection{DFT with the Dirac equation for uranium}
\begin{verbatim}
 Total energy:
               E           E_ref     error
 -28001.13232549 -28001.13232549  2.47E-09

 Eigenvalues:
   n               E           E_ref     error
   1  -4223.41902046  -4223.41902046  5.20E-09
   2   -789.48978233   -789.48978233  3.31E-10
   3   -761.37447597   -761.37447597  4.48E-10
   4   -622.84809456   -622.84809456  4.34E-10
   5   -199.42980564   -199.42980564  2.78E-10
   6   -186.66371312   -186.66371312  4.58E-10
   7   -154.70102667   -154.70102667  5.59E-10
   8   -134.54118029   -134.54118029  5.01E-10
   9   -128.01665738   -128.01665738  4.67E-10
  10    -50.78894806    -50.78894806  4.61E-10
  11    -45.03717129    -45.03717129  4.90E-10
  12    -36.68861049    -36.68861049  5.42E-10
  13    -27.52930624    -27.52930624  5.68E-10
  14    -25.98542891    -25.98542891  4.89E-10
  15    -13.88951423    -13.88951423  5.00E-10
  16    -13.48546969    -13.48546969  5.56E-10
  17    -11.29558710    -11.29558710  5.65E-10
  18     -9.05796425     -9.05796425  5.41E-10
  19     -7.06929563     -7.06929563  5.24E-10
  20     -3.79741623     -3.79741623  5.96E-10
  21     -3.50121718     -3.50121718  5.47E-10
  22     -0.14678838     -0.14678838  5.73E-10
  23     -0.11604716     -0.11604717  6.20E-10
  24     -1.74803995     -1.74803995  7.11E-10
  25     -1.10111900     -1.10111900  6.52E-10
  26     -0.77578418     -0.77578418  6.41E-10
  27     -0.10304081     -0.10304082  5.51E-10
  28     -0.08480202     -0.08480202  5.48E-10
  29     -0.16094728     -0.16094728  3.25E-10
\end{verbatim}

\end{document}